\documentclass[aps, prl, tightenlines, floatfix, twocolumn, amsmath, superscriptaddress, showpacs, showkeys]{revtex4-1}
\usepackage{graphicx}% Include figure files
\usepackage{dcolumn}% Align table columns on decimal point
\usepackage{bm}% bold math
\usepackage{float}
\usepackage{hyperref}
\usepackage{bigints}
\usepackage{xcolor}

\begin{document}

\title{Observability of the very-high-energy emission from GRB 221009A}

\author{Giorgio Galanti}
\email{gam.galanti@gmail.com}
\affiliation{INAF, Istituto di Astrofisica Spaziale e Fisica Cosmica di Milano, Via Alfonso Corti 12, I -- 20133 Milano, Italy}

\author{Lara Nava}
\email{lara.nava@inaf.it}
\affiliation{INAF, Osservatorio Astronomico di Brera, Via Emilio Bianchi 46, I -- 23807 Merate, Italy}
\affiliation{INFN, Sezione di Trieste, Via Alfonso Valerio 2, I -- 34127 Trieste, Italy}

\author{Marco Roncadelli}
\email{marcoroncadelli@gmail.com}
\affiliation{INFN, Sezione di Pavia, Via Agostino Bassi 6, I -- 27100 Pavia, Italy}
\affiliation{INAF, Osservatorio Astronomico di Brera, Via Emilio Bianchi 46, I -- 23807 Merate, Italy}

\author{Fabrizio Tavecchio}
\email{fabrizio.tavecchio@inaf.it}
\affiliation{INAF, Osservatorio Astronomico di Brera, Via Emilio Bianchi 46, I -- 23807 Merate, Italy}

\author{Giacomo Bonnoli}
\email{giacomo.bonnoli@inaf.it}
\affiliation{INAF, Osservatorio Astronomico di Brera, Via Emilio Bianchi 46, I -- 23807 Merate, Italy}

\date{\today}
\begin{abstract}
The LHAASO Collaboration detected the gamma ray burst GRB 221009A at energies above $500 \, {\rm GeV}$ with a tail extending up to $18 \, \rm TeV$, whose spectral analysis has presently been performed up to $7 \, \rm TeV$ for the lower energy instrument LHAASO-WCDA only, with no indication of a cutoff. Soon thereafter, Carpet-2 at Baksan Neutrino Observatory reported the observation of an air shower consistent with being caused by a photon of energy $251 \, {\rm TeV}$ from the same GRB. Given the source redshift $z=0.151$, the expected attenuation due to the extragalactic background light is very severe so that these detections have proven very hard to explain. In this Letter, we show that the existence of axion-like-particles (ALPs) with mass $m_a \simeq (10^{-11}-10^{-7}) \, {\rm eV}$ and two-photon coupling $g_{a \gamma \gamma} \simeq (3-5) \times 10^{-12} \, {\rm GeV}^{- 1}$ strongly reduce the optical depth of TeV photons, thus explaining the observations. Our ALPs meet all available constraints, are consistent with  two previous hints at their existence and are good candidates for cold dark matter. Moreover, we show that Lorentz Invariance Violation (LIV) can explain the Carpet-2 result but {\it not} the LHAASO observations.
\end{abstract}

\maketitle

%%%%%%%%%%%%%%%%%%%%%%%%%%%%%%%%%%%%%%%%%%%%
%% MAINMATTER
%%%%%%%%%%%%%%%%%%%%%%%%%%%%%%%%%%%%%%%%%%%%

\noindent {\it Introduction} -- On 2022/10/11, the LHAASO Collaboration reported the detection of more than 5000 very-high-energy (VHE, ${\cal E} > 100 \, {\rm GeV}$) photons from the Gamma Ray Burst GRB 221009A. Specifically, GRB 221009A was detected by LHAASO-WCDA above $500 \, \rm GeV$ within 2000 seconds after the initial burst with significance over 100$\sigma$, and by LHAASO-KM2A with significance about 10$\sigma$ with the highest photon energy reaching $18 \, \rm TeV$~\cite{LHAASO}. Absorption features in the optical spectrum allowed to measure the redshift, which is $z = 0.151$~\cite{redshift}.
Before this discovery, the largest photon energy ever detected from a GRB was about 3\,TeV observed by H.E.S.S. from GRB 190829A at $z=0.079$~\cite{hess}.

On 2022/10/12, the Carpet-2 Collaboration at Baksan Neutrino Observatory announced the detection of an air shower consistent with being caused by a photon of energy $251 \, {\rm TeV}$ from a position encompassing GRB 221009A and observed 4536 seconds after the Fermi-GBM trigger~\cite{carpet}.  

The detection of photons with energies ${\cal E} > 10 \, \rm TeV$ is a {\it very important  discovery}, as the flux of VHE photons from extragalactic sources is strongly attenuated through $e^+ e^-$ pair production on the extragalactic background light (EBL). As we shall see, at $z = 0.151$ the attenuation of a photon flux with ${\cal E} = 18 \, {\rm TeV}$ is at least ${\cal O} (10^6 - 10^8)$, making extremely difficult to explain the LHAASO and even more the Carpet-2 detection (below we shall pay more attention to the LHAASO events  because of their higher reliability). One possibility to solve the problem is to invoke unconventional particle physics, such as axion-like particles (ALPs), a scenario which we have been deeply investigating over the last several years~\cite{universe}.

In this Letter we propose -- for the first time -- that ALPs explain the detection of multi-TeV photons from any GRB, and specifically from GRB 221009A~\cite{altri}. As we shall show, the gist of our strategy is that beam photons can convert into ALPs whose propagation is unaffected by the EBL, so that the effective optical depth $\tau_{\rm ALP} ({\cal E})$ is strongly reduced. Furthermore, we analyze the implications at ${\cal E} >10 \, \rm TeV$ of the spectral data of GRB 221009A released so far by the LHAASO Collaboration, which are limited to the low energy instrument [LHAASO-WCDA, $(0.2-7) \, \rm TeV$] and show consistency with a power-law spectrum up to $\sim 7 \, \rm TeV$ with no indication of a cutoff~\cite{grb221009aSpectrum} (see also Appendix). We also show that Lorentz Invariance Violation (LIV) can explain the Carpet-2 result but {\it not} the LHAASO observations.

\vspace{3mm}

\noindent {\it Cosmic photon extinction} -- On their way to us from extragalactic sources, VHE photons scatter off the EBL, namely the infrared-optical-ultraviolet photon background emitted by the whole stellar population and possibly reprocessed by dust during the cosmic evolution (for a review, see~\cite{dwek}). Several EBL models have been proposed so far  and they are discussed in the Supplemental Material (SM~\cite{SMref1}). Given an EBL model, the resulting optical depth $\tau_{\rm CP} ({\cal E})$ can be computed, in terms of which the photon survival probability reads $P_{\rm CP} ({\cal E}; \gamma \to \gamma) = e^{- \tau_{\rm CP}} ({\cal E})$. Here we employ the Saldana-Lopez {\it et al.} EBL model~\cite{saldanalopez}  -- to be referred to as SL EBL -- because of four different reasons: 1) it is the most recent one; 2) it is based on the deepest galaxy data sets ever obtained; 3) it is derived by a satellite borne detector, which minimizes foreground effects (like zodiacal light); 4) it is the one used by the LHAASO  Collaboration to infer their spectral results~\cite{grb221009aSpectrum}. The corresponding values of the photon survival probability for the nominal photon energies reported by LHAASO and Carpet-2 -- and also at lower energies to allow for an uncertainty of (15 - 20) \%~\cite{astri-lhaaso} and  50 \%, respectively --  are: $P_{\rm CP} (15 \, {\rm TeV} ; \gamma \to \gamma) = 3 \times 10^{- 6}$, $P_{\rm CP} (18 \, {\rm TeV} ; \gamma \to \gamma) = 1 \times 10^{- 8}$, $P_{\rm CP} (100 \, {\rm TeV} ; \gamma \to \gamma) = 3 \times 10^{- 96}$ and $P_{\rm CP} (251 \, {\rm TeV} ; \gamma \to \gamma) \sim 0$ (see also~\cite{manuelGRB}). Thus, within conventional physics alone, ${\cal E} > 10 \, {\rm TeV}$ photons are extremely unlikely to detect since their observation would require a huge TeV luminosity which is in tension with model predictions (see last Section).

\vspace{3mm}

\noindent {\it Axion-like particles (ALPs)} -- Many extensions of the Standard Model of particle physics -- especially superstring and superbrane theories -- predict the existence of ALPs (see e.g.~\cite{turok1996,string1,string2,string3,string4,string5,axiverse,abk2010,cicoli2012,dias2014,scott2017} and references therein, and~\cite{JR2010,r2012,irastorzaredondo,universe}  for reviews). Here, we provide only the minimal information about ALPs needed to understand the proposed scenario (details are reported in SM~\cite{SMref2}). 

ALPs are very light pseudo-scalar bosons of mass $m_a$ and with two-photon coupling $g_{a \gamma \gamma}$ described by the Lagrangian 
\begin{equation}
{\cal L}_{a \gamma \gamma} = - \, \frac{1}{4} \, g_{a \gamma \gamma} \, F_{\mu \nu} \, {\tilde{F}}^{\mu \nu} \, a = g_{a \gamma \gamma} \, {\bf E} \cdot {\bf B} \, a~,
\label{a2}
\end{equation}
where $a$ is the ALP field and $F_{\mu \nu}$ is the electromagnetic tensor with electric and magnetic components ${\bf E}$ and ${\bf B}$, respectively, and ${\tilde{F}}^{\mu \nu}$ is its dual (in Eq.~(\ref{a2}) ${\bf E}$ is the photon electric field). In this context two additional effects should be considered, namely the QED vacuum polarization~\cite{hew1,hew2,hew3,rs} and the photon dispersion on the CMB~\cite{raffelteffect} (see SM). Other couplings to conventional particles can exist but are irrelevant here, hence they are discarded. Actually, Eq.~(\ref{a2}) entails that in the presence of an external magnetic field ${\bf B}$ $\gamma \to a$ and $a \to \gamma$ {\it conversions} take place, thereby implying that as a photon beam propagates these conversions produce $\gamma \leftrightarrow a$ {\it oscillations}~\cite{mpz,rs}. Still, within this scenario ALPs do not effectively interact either with single photons or with matter~\cite{grjhea}. As a consequence  
$\tau_{\rm ALP} ({\cal E})$ gets reduced, but -- since the photon survival probability is presently $P_{\rm ALP} ({\cal E}; \gamma \to \gamma) = e^{- \, \tau_{\rm ALP}(\cal E)}$ -- even a small decrease of $\tau_{\rm ALP} ({\cal E})$ with respect to $\tau_{\rm CP} ({\cal E})$ gives rise to a large increase of $P_{\rm ALP} ({\cal E}; \gamma \to \gamma)$~\cite{drm2007} (see SM). 

The photon-ALP interaction gives rise to several effects both on the observed spectra (with possible hints at the ALP existence, see e.g.~\cite{drm2007,simet2008,sanchezconde2009,dgr2011,trgb2012,wb2012,trg2015,kohri2017,gtre2019,grdb}) and on the final photon polarization (see e.g.~\cite{ALPpol1,bassan,ALPpol2,ALPpol3,ALPpol5,day,galantiTheo,galantiPol,grtcClu,grtBlazar}). 
Under the assumption ${\cal E} \gg m_a$ -- which is presently satisfied -- the beam propagation equation obeys a Schr\"odinger-like equation with time replaced with the beam propagation direction $y$, so that a photon/ALP beam can be treated as a non-relativistic quantum system~\cite{rs}. This fact entails that the pure states of the beam are described by the wave function $\psi (y) = \bigl(A_x (y), A_z (y), a (y) \bigr)$ with $A_x (y)$, $A_z (y)$, $a (y)$ denoting the photon and ALP amplitudes, respectively (see SM). Moreover, the {\it transfer matrix} of the beam over the distance interval $[y_i, y_{i + 1}]$ is denoted by ${\cal U}_i ({\cal E}; y_{i + 1}, y_i)$, and the total transfer matrix over a number $N$ of intervals is 
\begin{equation}
{\cal U} ({\cal E}; y_{N + 1}, y_1) = \prod_{i = 1}^N {\cal U}_i ({\cal E}; y_{i + 1}, y_i)~. 
\label{14022023q}
\end{equation}
Assuming unpolarized emitted photons, we must employ the polarization density matrix $\rho (y) \equiv |\psi (y) \rangle \langle \psi (y)|$. 
Therefore, the overall photon survival probability is
\begin{eqnarray}
&\displaystyle P_{\rm ALP} ({\cal E}; \gamma \to \gamma) = \sum_{i = x,z} {\rm Tr}  \bigl[\rho_i \, {\cal U} ({\cal E}; y_{N+1}, y_1) \times \nonumber    \\
&\displaystyle \times \rho_{\rm unp} \, {\cal U}^{\dagger} ({\cal E}; y_{N+1}, y_1) \bigr]~,       \label{q2}
\end{eqnarray}
where $\rho_x \equiv {\rm diag} (1, 0, 0)$, $\rho_z \equiv {\rm diag} (0, 1, 0)$ and 
$\rho_{\rm unp} \equiv {\rm diag} (0.5, 0.5, 0)$~\cite{dgr2011}. The quantities $m_a$ and $g_{a \gamma \gamma}$ are totally unrelated, and we take $10^{- 12} \, {\rm eV} \leq m_a \leq 10^{-6} \, {\rm eV}$ and $10^{- 13} \, {\rm GeV}^{- 1} \leq g_{a \gamma \gamma} \leq 10^{- 10} \, {\rm GeV}^{- 1}$ as starting ALP parameter space. 

\vspace{3mm}

\noindent {\it ALP scenario applied to GRB 221009A} -- In order to enhance the clarity of our investigation, we consider at once the individual media crossed by the photon/ALP beam as it travels from the source to us. Below, we outline the logic of our calculations, which are reported in great detail in SM.

\vspace{2mm}

\noindent 1) {\it Conversion inside the source:} The exact time when photons above 10\,TeV have been recorded is currently unknown, but the timescale (within 2000 s) suggests that this detection is related to the afterglow emission, similarly to all previous detections of TeV GRBs~\cite{magic1,magic2,hess}. In this scenario, photons are produced in the downstream region of the  forward shock. The path traveled by the photons in the shocked region is of order $R/\Gamma$ (comoving frame), where $R$ is the distance from the central engine and $\Gamma$ the bulk Lorentz factor. We take $R\sim2\times10^{17}$\,cm, $\Gamma=45$, comoving electron density $n'=450$\,cm$^{-3}$ and comoving magnetic field strength $B'=2$\,G (see SM for justification~\cite{SMref3}). Given these values and by following a similar procedure as in~\cite{gtre2019}, we compute the transfer matrix ${\cal U}_1 ({\cal E}; y_2, y_1)$ in the GRB jet, where $y_2$ and $y_1$ denote the position of the border of the GRB and of the production region, respectively. We find that photon-ALP conversions in the source are negligible, whence ${\cal U}_1 ({\cal E}; y_2, y_1) \simeq 1$.

\vspace{2mm}

\noindent 2) {\it Conversion inside the host galaxy:} In~\cite{GRB221009Ahost} evidence that the host galaxy of GRB~221009A is a disc-like one is presented. In the lack of any firm knowledge of its nature, we consider the two most likely possibilities: a typical spiral (see e.g.~\cite{SpiralBrev,Fletcher2010}) and a starburst with intermediate properties similar to 
M82~\cite{Thompson2006,LopezRodriguez2021}. According to~\cite{GRB221009Ahost}, the host galaxy has an edge-on orientation -- hence causing the photon/ALP beam to propagate inside the disk -- with the GRB located close to the nuclear region.  So, we place our GRB in the neighborhood of the galactic center (see also~\cite{GRBposition,GRBposition2}). We take into account all components of the host magnetic field ${\bf B}_{\rm host}$ with their stochastic properties~\cite{SpiralBrev,Elmegreen2004,Haverkorn2008} and their radial profiles~\cite{Heesen2023}, which are relevant for the photon-ALP conversion. The magnetic field strength assumed for the spiral is $B_{\rm host, spiral} = {\cal O}(5-10) \, \mu{\rm G}$, while for the starburst is $B_{\rm host, starburst} = {\cal O}(20-50) \, \mu{\rm G}$. Knowing the behavior of ${\bf B}_{\rm host}$ and of the other relevant parameters for either a spiral or a starburst hosting galaxy, we compute the transfer matrix in the host ${\cal U}_2 ({\cal E}; y_3, y_2)$, where $y_3$ denotes the position of the external luminous galaxy edge.

\vspace{2mm}

\noindent 3) {\it Conversion in extragalactic space:} Unfortunately, our knowledge of the extragalactic magnetic field ${\bf B}_{\rm ext}$ is still very poor. All we know is that $B_{\rm ext}$ lies in the range $10^{- 7} \, {\rm nG} \lesssim B_{\rm ext} \lesssim 1.7 \, {\rm nG}$ on the scale of ${\cal O} (1) \, {\rm Mpc}$~\cite{neronovvovk,durrerneronov,upbbext}. Nevertheless, it has become customary to model $B_{\rm ext}$ as a domain-like network, wherein ${\bf B}_{\rm ext}$ is assumed to be homogeneous over a whole domain of size $L_{\rm dom}^{\rm ext}$ equal to its coherence length, with ${\bf B}_{\rm ext}$ changing randomly its direction from one domain to the next, keeping approximately the same strength~\cite{kronberg1994,grassorubinstein2001}. Accordingly, the beam propagation becomes a {\it random process}, and only a single realization at once can be observed. We employ a recent and physically accurate model to describe ${\bf B}_{\rm ext}$ with $B_{\rm ext}= {\cal O} (1) \, {\rm nG}$ and $L_{\rm dom}^{\rm ext} = {\cal O} (1) \, {\rm Mpc}$~\cite{galantironcadelli20118prd,kartavtsev}. The latter possibility -- our option 1 -- is suggested by several scenarios~\cite{reessetti1968,hoyle1969,kronbergleschhopp1999,furlanettoloeb2001}.  Given the above uncertainty, we also consider the very conservative value $B_{\rm ext} < 10^{- 15} \, {\rm G}$ (option 2). In either case, we take the photon dispersion on the CMB  into account. Having fixed the properties of ${\bf B}_{\rm ext}$ and following~\cite{grjhea} we evaluate the transfer matrix ${\cal U}_3 ({\cal E}; y_4, y_3)$ in the extragalactic space for both options, where $y_4$ denotes the position of the outer luminous edge of the Milky Way. 

\vspace{2mm}

\noindent 4) {\it Conversion in the Milky Way:} The morphology of the magnetic field ${\bf B}_{\rm MW}$ and of the electron number density $n_{{\rm MW}, e}$ in the Galaxy are nowadays rather well known. Concerning ${\bf B}_{\rm MW}$, we adopt the model by Jansson and Farrar~\cite{jansonfarrar1,jansonfarrar2,BMWturb}, which is more complete with respect to the one of Pshirkov {\it et al.}~\cite{pshirkov2011}, even though we do not find substantial differences by employing the latter. Regarding $n_{{\rm MW}, e}$, we use the model developed in~\cite{yaomanchesterwang 2017}. The transfer matrix in the Milky Way ${\cal U}_4 ({\cal E}; y_5, y_4)$ -- where $y_5$ is the position of the Earth -- is evaluated as in Section 3.4 of~\cite{gtre2019}. 

\vspace{3mm}

\begin{figure}
\begin{center}
\includegraphics[width=.48\textwidth]{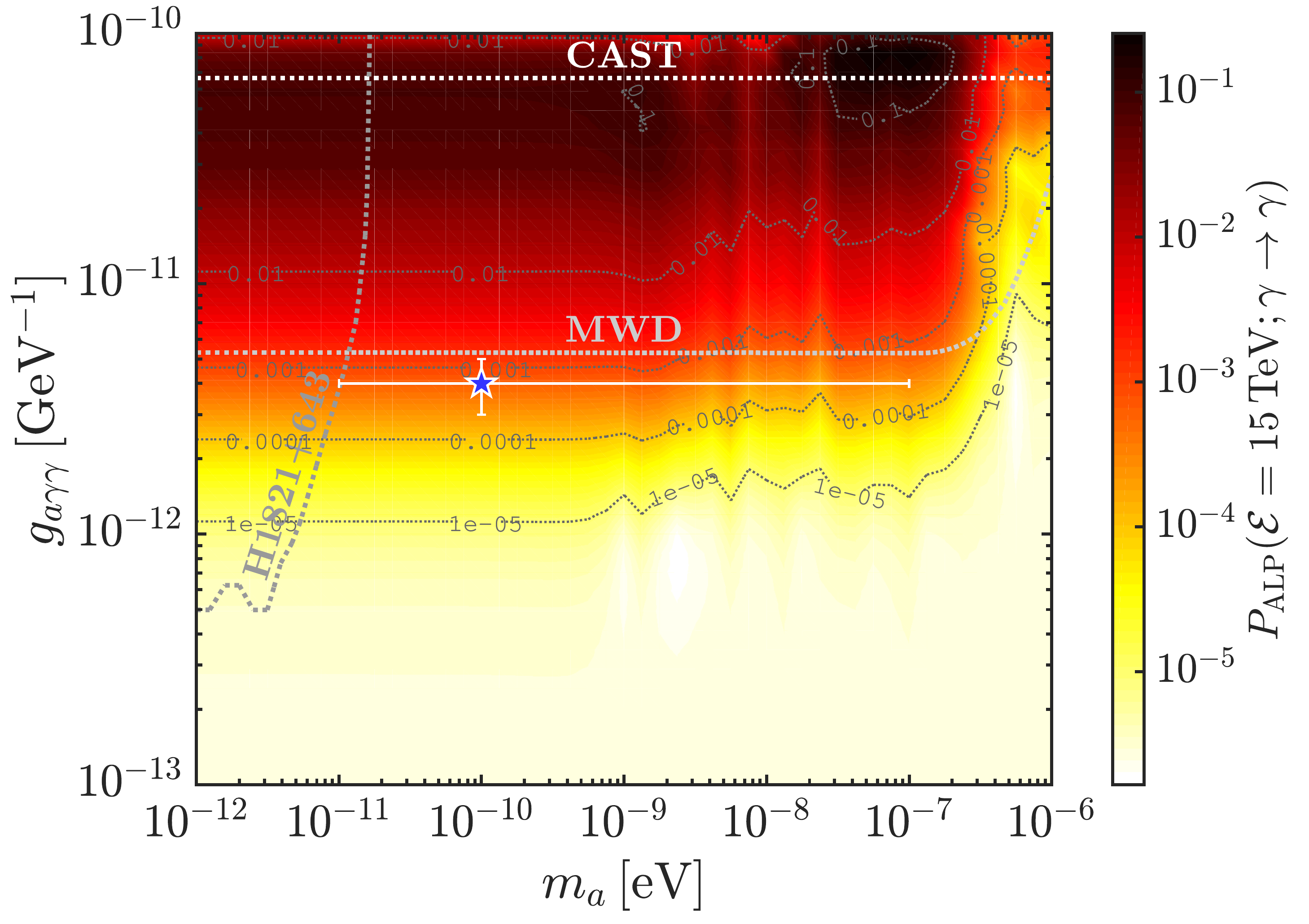}
\end{center}
\caption{\label{parSpaceStarburst} Photon survival probability $P_{\rm ALP}({\cal E}; \gamma \to \gamma)$ at energy ${\cal E} = 15 \, \rm TeV$ as a function of the ALP mass $m_a$ and of the photon-ALP coupling $g_{a\gamma\gamma}$ assuming the SL EBL model, $B_{\rm ext}=1 \, \rm nG$ and a starburst hosting galaxy. The blue star with error bar represents our choice for the ALP parameters: $m_a=10^{-10} \, \rm eV$ and $g_{a\gamma\gamma}=4 \times 10^{-12} \, \rm GeV^{-1}$. The three bounds mentioned in the text are also plotted.}
\end{figure}

\noindent {\it Results within the ALP model} -- Once all individual transfer matrices are known, the total one from the photon production region in GRB 221009A to the Earth is given by Eq.~(\ref{14022023q}) and the photon survival probability in the presence of photon-ALP interaction $P_{\rm ALP} ({\cal E}; \gamma \to \gamma)$ follows from Eq.~(\ref{q2}). Fig.~\ref{parSpaceStarburst} shows $P_{\rm ALP} ({\cal E}; \gamma \to \gamma)$ at ${\cal E} = 15 \, \rm TeV$ for values of $m_a$ and $g_{a\gamma\gamma}$ in the range $10^{- 12} \, {\rm eV} \leq m_a \leq 10^{- 6} \, {\rm eV}$ and $10^{- 13} \, {\rm GeV}^{- 1} \leq g_{a \gamma \gamma} \leq 10^{- 10} \, {\rm GeV}^{- 1}$ (see SM for similar plots at different energies). We assume $B_{\rm ext} = 1 \, \rm nG$ and a starburst hosting galaxy. 

Our next step concerns the ALP parameter space. To date the most reliable bounds are as follows: $g_{a\gamma\gamma} < 0.66 \times 10^{- 10} \, {\rm GeV}^{- 1}$ for $m_a < 0.02 \, {\rm eV}$ from the CAST experiment~\cite{cast}, $g_{a\gamma\gamma} < 6.3 \times 10^{- 13} \, {\rm GeV}^{- 1}$ for $m_a < 10^{- 12} \, {\rm eV}$ from observations of H1821+643  in the X-ray band~\cite{limJulia} and $g_{a\gamma\gamma} < 5.4 \times 10^{- 12} \, {\rm GeV}^{- 1}$ for $m_a < 3 \times 10^{- 7} \, {\rm eV}$ from the polarimetric analysis of magnetic white dwarfs (MWD)~\cite{mwd}. Within these bounds, the values of $m_a$ and $g_{a\gamma\gamma}$ which maximize $P_{\rm ALP}({\cal E}=15 \rm \,TeV; \gamma \to \gamma)$ are $m_a \simeq (10^{-11}-10^{-7}) \, \rm eV$ and $g_{a\gamma\gamma} \simeq (3-5) \times 10^{-12} \, \rm GeV^{-1}$. We choose $m_a=10^{-10} \, \rm eV$ and $g_{a\gamma\gamma} = 4 \times 10^{-12} \, \rm GeV^{-1}$ as benchmark values based on two previous hints at the ALP existence~\cite{trgb2012,grdb}. We report both $P_{\rm CP} ({\cal E}; \gamma \to \gamma)$ and $P_{\rm ALP} ({\cal E}; \gamma \to \gamma)$ in Fig.~\ref{survProbFigStarburst} for the case of a starburst galaxy hosting the GRB. 
\begin{figure}
\begin{center}
\includegraphics[width=.45\textwidth]{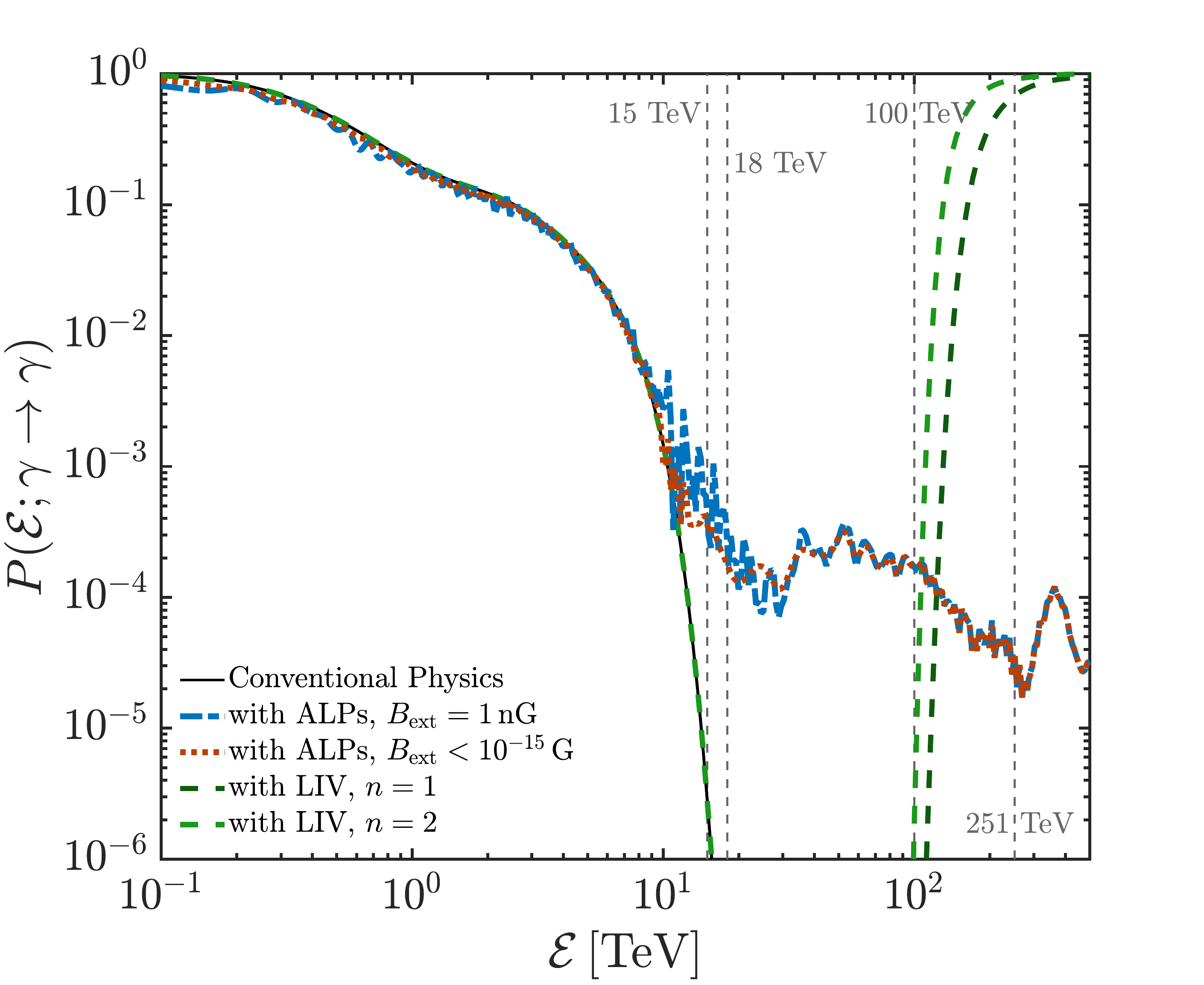}
\end{center}
\caption{\label{survProbFigStarburst} Photon survival probability $P ({\cal E}; \gamma \to \gamma)$ versus energy ${\cal E}$ in conventional physics, taking into account LIV and ALP effects in the case of a starburst hosting galaxy. We assume $m_a=10^{-10} \, \rm eV$ and $g_{a\gamma\gamma}=4 \times 10^{-12} \, \rm GeV^{-1}$.}
\end{figure}      
As discussed in~\cite{grjhea}, above ${\cal O}(5) \, \rm TeV$ the photon-ALP conversion in the extragalactic space becomes inefficient because of the photon dispersion on the CMB.  In Fig.~\ref{survProbFigStarburst} we consider both the case of an efficient ($B_{\rm ext} = 1 \, \rm nG$) and negligible ($B_{\rm ext} < 10^{-15} \, \rm G$) photon-ALP conversion in the extragalactic space. Remarkably, even though the effect is reduced for $B_{\rm ext} < 10^{-15} \, {\rm G}$, we can still consistently explain the photon observations of {\it both} LHAASO {\it and} Carpet-2 with no substantial difference. Note that as $m_a$ approaches $m_a = {\cal O}(10^{-7}) \, \rm eV$ the conversion in extragalactic space becomes inefficient and we recover the case $B_{\rm ext} < 10^{-15} \, \rm G$, which produces similar results (see Figs.~\ref{parSpaceStarburst} and~\ref{survProbFigStarburst}). 

The recent LHAASO spectral data analysis up to $7 \, \rm TeV$ shows the spectrum of GRB 221009A in different time slices during the 2000 s observational time~\cite{grb221009aSpectrum}. In order to evaluate its detectability at higher energies, we compute its time averaged spectrum extended up to 
$18 \, {\rm TeV}$ as discussed in the Appendix. Hence, we exhibit in Fig.~\ref{SpectrumStarburstFig} the intrinsic averaged spectrum and the observed one both in conventional physics and when photon-ALP effects  are at work, along with the LHAASO sensitivity (see Appendix). Fig.~\ref{SpectrumStarburstFig} shows that conventional physics {\it cannot} account for the detection of photons with ${\cal E} > 10 \, \rm TeV$ even in the best scenario wherein the spectrum continues as a power law, and in particular with energies even below the lower limit of the uncertainty about the LHAASO event at $18 \, \rm TeV$, while our ALP scenario with $m_a \simeq (10^{-11}-10^{-7}) \, \rm eV$ and $g_{a\gamma\gamma} \simeq (3-5) \times 10^{-12} \, \rm GeV^{-1}$ explains the LHAASO observations (see Appendix).

\begin{figure}
\begin{center}
\includegraphics[width=.45\textwidth]{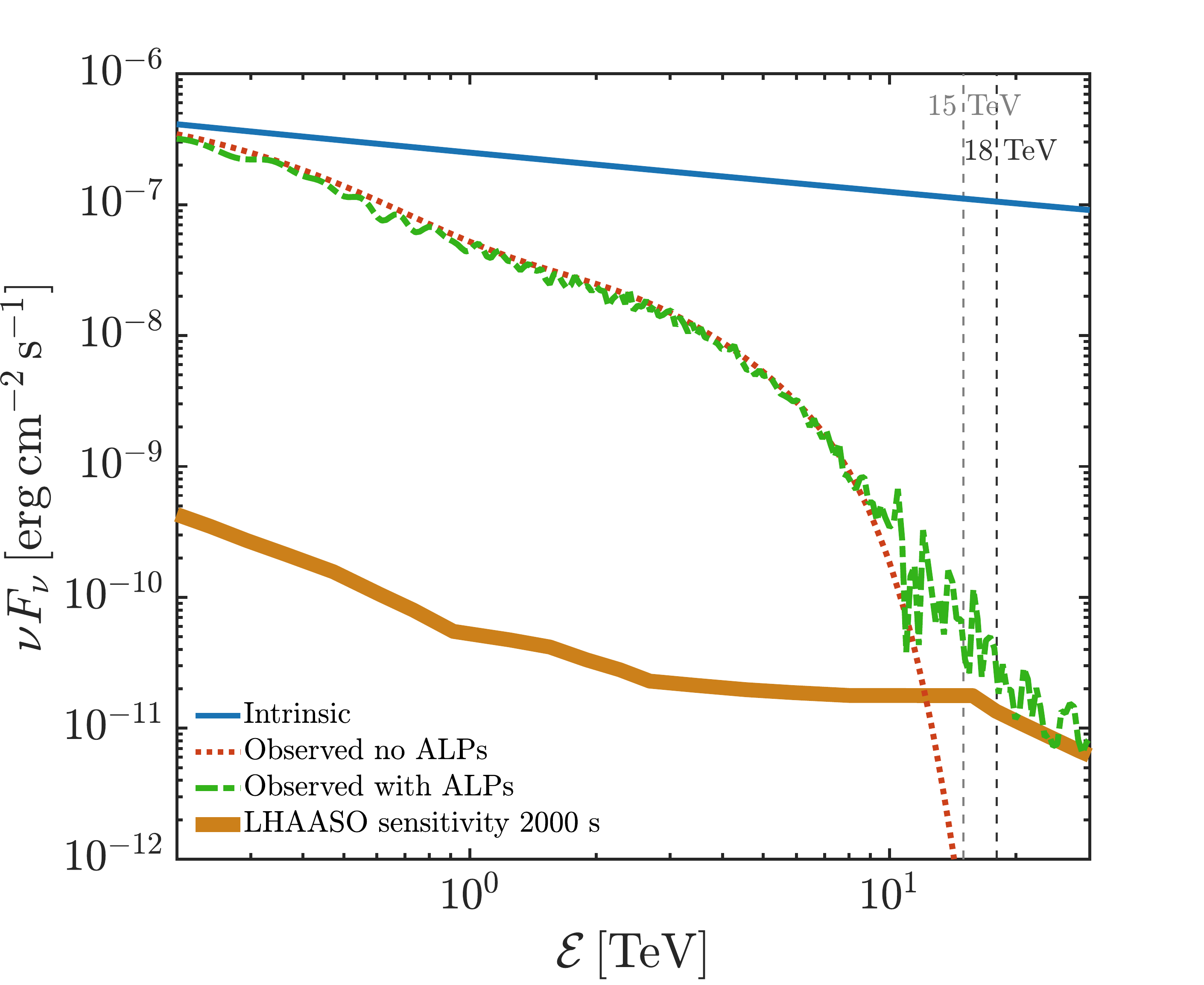}
\end{center}
\caption{\label{SpectrumStarburstFig} Intrinsic average spectrum of GRB 221009A as measured by LHAASO~\cite{grb221009aSpectrum} extended up to $\sim 20 \, \rm TeV$ and corresponding observed one versus energy ${\cal E}$ within conventional physics and when ALP effects are taken into account.  The LHAASO sensitivity at $2000 \, \rm s$ is also shown (see Appendix).}
\end{figure}   

The case of a spiral galaxy hosting the GRB is reported in SM, and leads to results in the same ballpark of those derived for a starburst galaxy even if slightly smaller.

%\ \

\vspace{3mm}

\noindent {\it Lorentz Invariance Violation (LIV)} -- Extensions of the standard model of particle physics encompassing quantum gravity predict a violation of Lorentz invariance (for a review, see~\cite{addazi}). As far as the present analysis is concerned, its main implication is the following modification of the photon dispersion relation
\begin{equation}
{\cal E}^2-p^2 = - \frac{{\cal E}^{n+2}}{{\cal E}_{{\rm LIV}}^n}~,
\label{liv}
\end{equation}
where ${\cal E}$ and $p$ are the photon energy and momentum, respectively, while ${\cal E}_{{\rm LIV}}$ is the high-energy scale above which LIV becomes important. As already demonstrated in~\cite{gtl,tavLIV} for a redshift very close to that of GRB 221009A, within the current LIV limits~\cite{LIVlim} LIV {\it cannot} explain a reduced opacity at LHAASO energies~\cite{LHAASO} (see also~\cite{manuelGRB} and Fig.~\ref{survProbFigStarburst}). On the contrary, at energies around the detection by Carpet-2~\cite{carpet}, the photon survival probability $P_{\rm LIV} ({\cal E})$ in the presence of LIV effects approaches the value $P_{\rm LIV} ({\cal E})=1$ (see Fig.~\ref{survProbFigStarburst}), for both the cases of $n=1$ with ${\cal E}_{{\rm LIV}, n=1} = 3 \times 10^{29} \, {\rm eV}$ and $n=2$ with ${\cal E}_{{\rm LIV}, n=2} = 5 \times 10^{21} \, {\rm eV}$ within the current LIV bounds~\cite{LIVlim}.

%\ \

\vspace{3mm}

\noindent {\it Discussion and Conclusions} -- We restate that within conventional physics alone ${\cal E} > 10 \, {\rm TeV}$ photons are extremely unlikely to detect since their observation would require a huge TeV luminosity which is in tension with model predictions. Specifically, among the theoretical models proposed to explain the origin of ${\cal E} > 10 \, {\rm TeV}$ photons in GRB~221009A there are synchrotron self Compton (SSC) radiation and secondary emission from ultra-high energy protons \cite{mirabal,gonzalez,das,zhao,sahu}. In all these studies, the predicted emission -- once EBL attenuation and instrument efficiency are considered -- is found to have a low chance to be detected, and only for contrived choices of the model parameters~\cite{gonzalez}, and/or peculiar choices on the location of the GRB~\cite{mirabal,das}. Thus, observations appear to be in tension with the possibility to produce a ${\cal E} > 10 \, \rm TeV$ emission luminous enough to be detected by LHAASO within conventional physics. 

A most effective way to achieve the same goal is to invoke photon-ALP oscillations, a mechanism which strongly reduces the opacity of the Universe to VHE photons. This  possibility has been investigated in this Letter for the first time concerning TeV GRBs in general and for GRB 221009A in particular. Taking uncertainties into account, we have shown that ALPs with mass $m_a \simeq (10^{-11} -10^{-7}) \, \rm eV$ and two-photon coupling $g_{a\gamma\gamma} \simeq (3-5) \times 10^{-12} \, \rm GeV^{-1}$ significantly increase the photon survival probability of multi-TeV photons and explain the LHAASO detection above $10 \, \rm TeV$ in conjunction with the SL EBL model, which is the best one available to date. We stress that the considered values of both $m_a$ and $g_{a \gamma \gamma}$ are in agreement with the strongest bounds. Moreover, these values are consistent with two previous hints at the ALP existence~\cite{trgb2012,grdb} and make them good candidates for cold dark matter~\cite{arias2012}. 

We also investigated the alternative LIV scenario, finding that it is {\it unable} to explain photon detection at LHAASO energies within current limits, but {\it can} explain that by Carpet-2. 

The ALP parameter space employed here will be probed with astrophysical data collected by several observatories such as ASTRI~\cite{astri}, CTA~\cite{cta}, GAMMA-400~\cite{g400}, HAWC~\cite{hawc}, HERD~\cite{herd}, LHAASO~\cite{LHAASOsens} and TAIGA-HiSCORE~\cite{desy}, by laboratory experiments like ALPS II~\cite{alps2}, IAXO~\cite{iaxo,iaxo2} and STAX~\cite{stax}, with the techniques developed by Avignone and collaborators~\cite{avignone1,avignone2,avignone3}, and possibly by the ABRACADABRA experiment~\cite{abracadabra}.

\ \

\noindent {\it Note added} -- After the acceptance of this Letter, the LHAASO Collaboration released a preprint on a first analysis of the GRB 221009A spectrum above $10 \, \rm TeV$~\cite{LHAASOspectrumHigh}. Therein, the highest energy bin can be at ${\cal E} \sim 13 \, \rm TeV$ or at ${\cal E} \sim 18 \, \rm TeV$, depending on the adopted fitting procedure. According to the LHAASO Collaboration ``Observations of photons up to $13 \, \rm TeV$ from a source with a measured redshift of $z = 0.151$ requires more transparency in intergalactic space than previously expected, in order to avoid an unusual pile-up at the end of the spectrum. Alternatively, one may invoke new physics such as Lorentz Invariance Violation (LIV) or assume an axion origin of very high energy (VHE) signals''. This LHAASO conclusion makes our results about ALPs solving the problem even more robust, while we stress that, as shown in detail in our paper, LIV actually provides no benefit in explaining this experimental evidence.

\ \

\begin{acknowledgments}

{\it Acknowledgments} -- We thank Felix Aharonian, Patrizia Caraveo, Elisabete de Gouveia Dal Pino, Giovanni Pareschi, Isabella Prandoni and Franco Vazza for discussions and useful information. The work of G. G. is supported by a contribution from the grant ASI-INAF 2015-023-R.1. The work of M. R. is supported by an INFN grant. This work was made possible also by the funding by the INAF Mini Grant `High-energy astrophysics and axion-like particles', PI: Giorgio Galanti. The work by L. N. is partially supported by the INAF Mini Grant `Shock acceleration in Gamma Ray Bursts', PI: Lara Nava.

\end{acknowledgments}

\ \

\noindent{\bf Appendix}

\ \

\noindent {\it GRB 221009A spectrum} -- The spectrum of GRB 221009A has recently been released by LHAASO for energies $0.2 \, {\rm TeV} \lesssim {\cal E} \lesssim 7 \, {\rm TeV}$~\cite{grb221009aSpectrum}. The LHAASO Collaboration fits the intrinsic spectrum ${\cal F}_i$ in the $i$-th time interval once data are EBL-deabsorbed by means of the `standard', `high' and `low' SL EBL models with the power law  
\begin{equation}
{\cal F}_i({\cal E}) \equiv \frac{{\rm d}N}{{\rm d}{\cal E}}=A \left( \frac{{\cal E}}{{\rm TeV}}\right)^{-\gamma}~, 
\label{app1}
\end{equation}
where $A$ is a normalization constant expressed in units of $10^{-8} \, {\rm TeV}^{-1} \, {\rm cm}^{-2} \, {\rm s}^{-1}$ and $\gamma$ is the spectral slope. Different values of $A$ and $\gamma$ are reported in Table S2 of~\cite{grb221009aSpectrum} for 5 time intervals covering the full time range $(231-2000) \, \rm s$ of LHAASO detection after the Fermi-GBM trigger. The LHAASO Collaboration does not find any curvature in the intrinsic spectrum up to $7 \, \rm TeV$. Therefore, an extrapolation of Eq.~(\ref{app1}) up to about 20 TeV looks justified.

In order to compare the observed spectrum of GRB 221009A with the LHAASO sensitivity (see below) we compute the time averaged intrinsic spectrum ${\cal F}_{\rm int}$ defined as 
\begin{equation}
{\cal F}_{\rm int} \equiv \langle {\cal F}_i \rangle = \frac{1}{T} \sum_{i=1}^5 \Delta t_i {\cal F}_i~, 
\label{app2}
\end{equation}
where $T=1769 \, \rm s$ is the total duration of the LHAASO detection, while $\Delta t_i$ is the duration of the $i$-th time interval, as reported in Table S2 of~\cite{grb221009aSpectrum}.

The observed spectrum ${\cal F}_{\rm obs}$ can be evaluated by means of 
\begin{equation}
{\cal F}_{\rm obs}({\cal E}) = P({\cal E}; \gamma \to \gamma) {\cal F}_{\rm int}({\cal E})~, 
\label{app3}
\end{equation}
where $P({\cal E}; \gamma \to \gamma)=P_{\rm CP}({\cal E}; \gamma \to \gamma)$ in the case of conventional physics and $P({\cal E}; \gamma \to \gamma)=P_{\rm ALP}({\cal E}; \gamma \to \gamma)$ when ALP effects are considered (see the main text and SM for its derivation). We recall that the spectral energy distribution (SED) reported in Fig.~\ref{SpectrumStarburstFig} is related to ${\cal F}_{\rm obs}$ by $\nu F_{\nu}={\cal E}^2 {\cal F}_{\rm obs}$.

The observed time-averaged spectrum integrated over the energy error range of the instrument around the event at ${\cal E}=18 \, \rm TeV$ with an exposure time of $3600 \, \rm s$ and a KM2A effective area of $\sim 10^{9} \, {\rm cm}^2$~\cite{LHAASOarea} shows that conventional physics is unable to justify such a detection, while the ALP scenario considered in this Letter accomplishes the task. 

We have checked the robustness of our results by considering the `standard', `high' and `low' SL EBL models, an efficient or negligible photon-ALP conversion in the extragalactic space, and the cases of both a spiral and a starburst galaxy hosting the GRB. Passing from a `high' to a `low' SL EBL model and/or from an efficient to a negligible photon-ALP conversion in the extragalactic space and/or from a starburst to a spiral host galaxy, we progressively need higher values of $g_{a\gamma\gamma}$ inside the range $(3-5) \times 10^{-12} \, \rm GeV^{-1}$ to explain the LHAASO detection above $10 \, \rm TeV$. The value of $m_a$ is less constrained: the upper limit $m_a < 10^{-7} \, \rm eV$ is necessary in order to have an efficient photon-ALP conversion, while the lower limit $m_a > 10^{-11} \, \rm eV$ is imposed by the ALP bound concerning observations of H1821+643 in the X-ray band~\cite{limJulia}.

\vspace{3mm}

\noindent {\it LHAASO sensitivity} -- The LHAASO observatory is composed of two instruments: LHAASO-WCDA more sensitive for ${\cal E} < {\cal O}(15) \, \rm TeV$ and LHAASO-KM2A more sensitive at higher energies. The sensitivity curve of the entire LHAASO observatory for an exposure time of $1 \, \rm yr$ is reported in~\cite{LHAASOsens}.

We rescale such a curve for our purposes at an exposure time of $2000 \, \rm s$ taking into consideration that the yearly sensitivity includes periods when the source is below the horizon or the effective area drops significantly enough to neglect the contribution. For the declination of the GRB, LHAASO observes the position within 40 degrees of zenith distance for $\sim \, 6 \, \rm h$ per day, and we consider this daily time effect for rescaling the sensitivity. This is motivated from the published observation of Crab Nebula with the partial KM2A Array, where especially below few tens of TeV the effective area is negligible above 40 degrees~\cite{LHAASOarea}.

\begin{widetext}

\vskip 1 cm

\section{SUPPLEMENTAL MATERIAL}

\medskip
\medskip

\section{Details on the extragalactic background light (EBL)}

The optical depth of the EBL has first been computed by Nishikov~\cite{nishikov1962}, Gould and Schreder~\cite{gouldschreder1967}, and Fazio and Stecker~\cite{faziostecker1970}. It reads
\begin{eqnarray}
&\displaystyle \tau_{\rm EBL} ({\cal E}_0, z_s) = \int_0^{z_s} dz ~ \frac{d l (z)}{d z} \int_{- 1}^1 d ({\rm cos} \, \theta) \, \frac{1 - \, {\rm cos} \, \theta}{2} \int_{\epsilon_{\rm thrs} ({\cal E} (z), \theta)}^{\infty} d \epsilon (z) ~  \times   \nonumber \\
&\displaystyle \times \, n_{\rm EBL} \bigl(\epsilon (z), z \bigr) \, \sigma \bigl({\cal E} (z), \epsilon (z), \theta \bigr)~,   \label{marco10062017a}
\end{eqnarray}
where $\sigma \bigl({\cal E} (z), \epsilon (z), \theta \bigr)$ is the Breit Wheeler cross section~\cite{breitwheeler,heitler}. Moreover,  $\epsilon_{\rm thrs} ({\cal E} (z), \theta)$ is the threshold energy for emitted photons of energy ${\cal E} = (1+z) {\cal E}_0$ at generic redshift $z$, ${\cal E}_0$ is their observed energy, $z_s$ is the source redshift and $\theta$ is the scattering angle, while the distance traveled by a photon per unit redshift at redshift $z$ is given by
\begin{equation}
\frac{d l (z)}{d z} = \frac{c}{H_0} \frac{1}{(1 + z) \, \left[\Omega_{\Lambda} + \Omega_M \, (1 + z )^3 \right]^{1/2}} 
\label{marco10062017b}
\end{equation}
with Hubble constant $H_0 \simeq 70 \, {\rm km} \, {\rm s}^{- 1} \, {\rm Mpc}^{- 1}$, whereas 
$\Omega_{\Lambda} \simeq 0.7$ and $\Omega_M \simeq 0.3$ represent the average cosmic density of matter and dark energy, respectively, in units of the critical density $\rho_{\rm cr} = 0.97 \times 10^{- 29} \, {\rm g} \, {\rm cm}^{- 3}$.

Determining the spectral number density $n_{\rm EBL} \bigl(\epsilon (z), z \bigr)$ of the EBL is very difficult. It is affected by large uncertainties, arising mainly from foreground effects. Among them is the zodiacal light, which is various orders of magnitude larger than the EBL itself~\cite{hauwserdwek2001}. For this reason the most reliable EBL determination is derived from satellite borne detectors (more about this, later). Below, we briefly summarize seven basic strategies which has been pursued.

\medskip

\noindent {\bf Forward evolution:} This is the most ambitious approach, since it relies upon first principles, namely from semi-analytic models of galaxy formation in order to predict the time evolution of the galaxy luminosity function~\cite{primack2001,primack2005,gilmore2009,gilmore2012,inoue2013}. 

\medskip

\noindent {\bf Backward evolution from the ground:} This starts from observations from the ground of the present galaxy population and extrapolates the galaxy luminosity function backward in time. Among others, this strategy has been followed by~\cite{frv2008,franceschinirodighiero}.

\medskip

\noindent {\bf Backward evolution from the sky:} The logic is the same as the previous one but observations are performed with telescopes in space. Satellite-based observations have the great advantage to get rid of foreground effects and in particular of the zodiacal light, since different directions can easily be probed, thereby allowing the zodiacal light to be subtracted. Two recent implementations of this strategy are due to the CIBER Collaboration~\cite{ciber} and Saldana-Lopez {\it et al.} to be referred to as SL EBL model~\cite{saldanalopez}. The latter work gives by now the most robust determination of the EBL, since it is based on a sample of 150000 galaxies in five HST fields, and traces the EBL evolutionary history from the first galaxies at $z = 6$ until the present. Because of this reason it is employed in the present Letter.

\medskip

\noindent {\bf Inferred evolution:} This models the EBL by using quantities like the star formation rate, the initial mass function, and the dust extinction as inferred from 
observations~\cite{kneiske2002,kneiske2004,finke2010}.

\medskip

\noindent {\bf Minimal EBL:} This relies upon the same strategy underlying the previous item but with the parameters tuned in order to reproduce the EBL lower limits from galaxy counts. Note that population III stars cannot be accommodate within this model~\cite{kneiskedole2010}. Moreover, a lower limit to the EBL level has previously been  derived from integrated galaxy counts~\cite{madaupozzetti2000}. 

\medskip

\noindent {\bf Observed evolution:} This method has the advantage to rely only upon observations by using a very rich sample of galaxies extending over the redshift range $0 \leq z \leq 1$~\cite{dominguez2011}.

\medskip

\noindent {\bf Compared observations:} This technique has been implemented in two ways. One consists in comparing observations of the EBL itself with blazar observations with 
Imaging Atmospheric Cherenkov Telescopes (IACTs) and deducing the EBL level from the VHE photon dimming~\cite{schroedter2005,aharonian2006,mazinraue2007,mazingoebel2007,
finkerazzaque2009}. The other starts from some gamma-ray observations of a given blazar below 100 GeV where EBL absorption is negligible and infers the EBL level by comparing the IACT observations of the same blazar with the source spectrum as extrapolated from the former observations~\cite{orrkrennrichdwek2011}. In the latter case the main assumption is that the emission mechanism is presumed to be known with great accuracy. Anyway the crucial unstated assumption is that photon propagation in the VHE band is governed by conventional physics. As a consequence, this method prevents -- just by definition -- the discovery of ALPs.

\medskip

The values of the optical depth $\tau_{\rm CP}$ according to conventional physics (CP) for various EBL models and at some relevant energies $\cal E$ are reported in Table~\ref{tabEBL} along with the corresponding photon survival probability $P_{\rm CP} ({\cal E}; \gamma \to \gamma) = e^{- \, \tau_{\rm CP}({\cal E})}$ (see also~\cite{manuelGRB}). 
%These values are listed for the nominal photon energies reported by LHAASO and Carpet-2, and also at lower energies, to allow for an uncertainty of $(15-20)\%$~\cite{astri-lhaaso} and $~\sim 50 \%$, respectively, in the photon energy reported in~\cite{LHAASO,carpet}.

\medskip

\begin{table}

\label{tabEBL}
\begin{tabular}{l|cc|cc|cc|cc}
\hline
\hline
\multicolumn{1}{c|}{EBL} &\multicolumn{2}{c|}{$15 \, \rm TeV$} &\multicolumn{2}{c|}{$18 \, \rm TeV$} &\multicolumn{2}{c|}{$100 \, \rm TeV$} &\multicolumn{2}{c}{$251 \, \rm TeV$} \\

\hline

& $\tau_{\rm CP}$ & $P_{\rm CP}$  & $\tau_{\rm CP}$ & $P_{\rm CP}$ & $\tau_{\rm CP}$ & $P_{\rm CP}$ & $\tau_{\rm CP}$ & $P_{\rm CP}$ \\
D  &  12.7 & 3$\times$$10^{-6}$ & 19.4 & 4$\times$$10^{-9}$ & 350 & 2$\times$$10^{-152}$ & 9654 & $\sim$0 \\
G  &  9.4 & 8$\times$$10^{-5}$ & 13.1 & 2$\times$$10^{-6}$ & 246 & 2$\times$$10^{-107}$ & 9502 & $\sim$0 \\
FR & 10.1 & 4$\times$$10^{- 5}$ & 14.1 & 7$\times$$10^{- 7}$ & 333 & 2$\times$$10^{-145}$ & 15411 & $\sim$0 \\
SL &  12.8 & 3$\times$$10^{- 6}$ & 18.3 & $10^{- 8}$ & 220 & 3$\times$$10^{- 96}$ & $>$9251 & $\sim$0 \\

\hline
\hline
\end{tabular}
\caption{Values of the optical depth $\tau_{\rm CP}$ and corresponding attenuation factor $P_{\rm CP}$ at different energies for the EBL models of Dom\'inguez {\it et al.} (D)~\cite{dominguez2011}, Gilmore {\it et al.} (G)~\cite{gilmore2012}, Franceschini and Rodighiero (FR)~\cite{franceschinirodighiero} and Saldana-Lopez {\it et al.} (SL)~\cite{saldanalopez}.}
\end{table}

\section{Formalism of axion-like particles (ALPs)}

For the reader's convenience, In this Section we provide a self-consistent discussion of the aspects of ALPs which are relevant for the main text.

The standard Lagrangian for the photon-ALP system is
\begin{equation}
\label{lagr}
{\cal L}_{\rm ALP} =  \frac{1}{2} \, \partial^{\mu} a \, \partial_{\mu} a - \frac{1}{2} \, m_a^2 \, a^2 - \, \frac{1}{4 } g_{a\gamma\gamma} \, F_{\mu\nu} \tilde{F}^{\mu\nu} a = \frac{1}{2} \, \partial^{\mu} a \, \partial_{\mu} a - \frac{1}{2} \, m_a^2 \, a^2 + g_{a \gamma \gamma} \, 
{\bf E} \cdot {\bf B}~a~,
\end{equation}
where $a$ represents the ALP field, $F_{\mu\nu}$ is the electromagnetic tensor, whose electric and magnetic components are ${\bf E}$ and ${\bf B}$, respectively, $\tilde{F}^{\mu\nu}$ is the dual of $F_{\mu\nu}$, $m_a$ denotes the ALP mass while $g_{a \gamma \gamma}$ the two-photon coupling. From the ${\bf E} \cdot {\bf B}~a$ term in Eq.~(\ref{lagr}), we see that $a$ couples only with the component ${\bf B}_T$ of the external magnetic field ${\bf B}$ which is transverse to the photon momentum $\bf k$ (whereas ${\bf E}$ is the electric field of the propagating photon, see also~\cite{dgr2011}). 

In the presence of a strong external magnetic field, photon one-loop vacuum polarization effects should be taken into account and are described by the Heisenberg-Euler-Weisskopf (HEW) effective Lagrangian
\begin{equation}
\label{HEW}
{\cal L}_{\rm HEW} = \frac{2 \alpha^2}{45 m_e^4} \, \left[ \left({\bf E}^2 - {\bf B}^2 \right)^2 + 7 \left({\bf E} \cdot {\bf B} \right)^2 \right]~,
\end{equation}
where $\alpha$ is the fine-structure constant and $m_e$ is the electron mass~\cite{hew1, hew2, hew3}. We stress that at variance with the axion~\cite{cgn1995}, $m_a$ and $g_{a \gamma \gamma}$ are {\it totally unrelated} parameters. 

Let consider a photon-ALP beam of energy ${\cal E}$ propagating along the $y$-direction in a magnetized medium described by the vector ${\bf B}$ entering Eq.~(\ref{lagr}), while 
${\bf E}$ pertains to a propagating photon. As shown long ago~\cite{rs}, in the regime ${\cal E} \gg m_a$ -- which is manifestly satisfied in the present Letter -- the beam propagation is governed by a Schr\"odinger-like equation of the form
\begin{equation}
\label{propeq} 
\left(i \, \frac{d}{d y} + {\cal E} +  {\cal M} ({\cal E},y) \right)  \psi(y)= 0~,
\end{equation}
where ${\cal M} ({\cal E},y)$ is the mixing matrix and the wave function is
\begin{equation}
\label{psi} 
\psi(y)=\left(\begin{array}{c}A_x (y) \\ A_z (y) \\ a (y) \end{array}\right)~,
\end{equation}
with $A_x (y)$, $A_z (y)$, $a (y)$ denoting the two photon linear polarization amplitudes along the $x$ and $z$ axis, respectively and $a (y)$ stands for the ALP amplitude. 
Letting ${\cal U}({\cal E};y,y_0)$ be the {\it transfer matrix} of Eq. (\ref{propeq}) -- namely the particular solution obeying the condition ${\cal U}({\cal E};y_0,y_0)=1$ -- a generic solution of that equation can be represented as
\begin{equation}
\label{psi2} 
\psi(y)={\cal U}({\cal E};y,y_0)\psi(y_0)~.
\end{equation}

Much in the same way as wave functions describe pure states in quantum mechanics, here they represent polarized beams, but in analogy with quantum mechanics an unpolarized beam is represented by the polarization density matrix $\rho(y)$ satisfying the Von Neumann-like equation associated with Eq.~(\ref{propeq})
\begin{equation}
\label{vneum}
i \frac{d \rho (y)}{d y} = \rho (y) \, {\cal M}^{\dag} ( {\cal E}, y) - {\cal M} ( {\cal E}, y) \, \rho (y)~,
\end{equation}
whose solutions can be expressed in terms of ${\cal U} \bigl( {\cal E}; y, y_0 \bigr)$ as
\begin{equation}
\label{unptrmatr}
\rho ( y ) = {\cal U} \bigl({\cal E}; y, y_0 \bigr) \, \rho_0 \, {\cal U}^{\dag} \bigl({\cal E}; y, y_0 \bigr)~.
\end{equation}
This is important for us, since we consider an unpolarized photon beam. Hence, the probability describing a beam in the initial state $\rho_0$ at position $y_0$ and in the final state $\rho$ at position $y$ is given by
\begin{equation}
\label{unpprob}
P_{\rho_0 \to \rho} ({\cal E},y) = {\rm Tr} \Bigl[\rho \, {\cal U} ({\cal E}; y, y_0) \, \rho_0 \, {\cal U}^{\dag} ({\cal E}; y, y_0) \Bigr]~,
\end{equation}
with ${\rm Tr} \, \rho_0 = {\rm Tr} \, \rho =1$~\cite{dgr2011}.
 
Let us next address the mixing matrix ${\cal M} ( {\cal E}, y)$ entering Eq.~(\ref{propeq}). By defining the angle $\phi$ between ${\bf B}_T$ and the $z$ axis, ${\cal M} ( {\cal E}, y)$ has the form 
\begin{equation}
\label{mixmat}
{\cal M} ({\cal E},y) \equiv \displaystyle \left(
\begin{array}{ccc}
\Delta_{xx} ({\cal E},y) & \Delta_{xz} ({\cal E},y) & \Delta_{a \gamma}(y) \, {\rm sin} \, \phi \\
\Delta_{zx} ({\cal E},y) & \Delta_{zz} ({\cal E},y) & \Delta_{a \gamma}(y) \, {\rm cos} \, \phi \\
\Delta_{a \gamma}(y) \, {\rm sin}  \, \phi & \Delta_{ a \gamma}(y) \, {\rm cos} \, \phi & \Delta_{a a} ({\cal E}) \\
\end{array}
\right)~.
\end{equation}
The various terms entering ${\cal M} ({\cal E},y)$ are as follows
\begin{equation}
\label{deltaxx}
\Delta_{xx} ({\cal E},y) \equiv \Delta_{\bot} ({\cal E},y) \, {\rm cos}^2 \, \phi + \Delta_{\parallel} ({\cal E},y) \, {\rm sin}^2 \, \phi~,
\end{equation}
\begin{equation}
\label{deltaxz}
\Delta_{xz} ({\cal E},y) = \Delta_{zx} ({\cal E},y) \equiv \left(\Delta_{\parallel} ({\cal E},y) - \Delta_{\bot} ({\cal E},y) \right) {\rm sin} \, \phi \, {\rm cos} \, \phi~,
\end{equation}
\begin{equation}
\label{deltazz}
\Delta_{zz} ({\cal E},y) \equiv \Delta_{\bot} ({\cal E},y) \, {\rm sin}^2 \, \phi + \Delta_{\parallel} ({\cal E},y) \, {\rm cos}^2 \, \phi~,
\end{equation}
\begin{equation}
\label{deltamix} 
\Delta_{a \gamma}(y) = \frac{1}{2}g_{a\gamma\gamma}B_T(y)~,
\end{equation}
\begin{equation}
\label{deltaM} 
\Delta_{aa} ({\cal E}) = - \frac{m_a^2}{2 {\cal E}}~,
\end{equation}
and
\begin{equation}
\Delta_{\bot} ({\cal E},y) = \frac{i}{2 \, \lambda_{\gamma} ({\cal E},y)} - \frac{\omega^2_{\rm pl}(y)}{2 {\cal E}} + \frac{2 \alpha}{45 \pi} \left(\frac{B_T(y)}{B_{{\rm cr}}} \right)^2 {\cal E} + \rho_{\rm CMB} \, {\cal E}~, 
\label{deltaort} 
\end{equation}
\begin{equation}
\Delta_{\parallel} ({\cal E},y) = \frac{i}{2 \, \lambda_{\gamma} ({\cal E},y)} - \frac{\omega^2_{\rm pl}(y)}{2 {\cal E}} + \frac{7 \alpha}{90 \pi} \left(\frac{B_T(y)}{B_{{\rm cr}}} \right)^2 {\cal E} + \rho_{\rm CMB} \, {\cal E} ~,  
\label{deltapar} 
\end{equation}
where $B_{{\rm cr}} \simeq 4.41 \times 10^{13} \, {\rm G}$ is the critical magnetic field and $\rho_{\rm CMB}=0.522 \times 10^{-42}$. Eq.~(\ref{deltamix}) accounts for the photon-ALP interaction, while Eq.~(\ref{deltaM}) describes the ALP mass effect. The first term in Eqs.~(\ref{deltaort}) and~(\ref{deltapar}) accounts for photon absorption due to the EBL -- as explained above we employ the SL EBL model~\cite{saldanalopez} -- and $\lambda_{\gamma}$ is the $\gamma\gamma \to e^+e^-$ mean free path~\cite{foot1}. In the second term of Eqs.~(\ref{deltaort}) and~(\ref{deltapar}) $\omega_{\rm pl}$ is the plasma frequency, which is related to the electron number density $n_e$ by $\omega_{\rm pl}=(4 \pi \alpha n_e / m_e)^{1/2}$. The third term in Eqs.~(\ref{deltaort}) and~(\ref{deltapar}) accounts for the photon one-loop vacuum polarization coming from ${\cal L}_{\rm HEW}$ of Eq.~(\ref{HEW}) -- which produces polarization variation and birifrangence on the beam -- while the fourth term represents the contribution from photon dispersion on the cosmic microwave background (CMB)~\cite{raffelteffect} which gives rise to sizable effects in the extragalactic space~\cite{grjhea}.

Thus, we see that for ${\cal E} \gg m_a$ the relativistic photon-ALP beam is formally described as a three-level non-relativistic unstable quantum system. Moreover, since the mixing matrix (\ref{mixmat}) is off-diagonal, the propagation eigenstates differ from the interaction eigenstates, thereby leading to $\gamma \leftrightarrow a$ oscillations quite  similar to oscillations of massive neutrinos with different flavors: the only difference is that in the present case an external ${\bf B}$ field is necessary in order to compensate for the spin mismatch between photons and ALPs~\cite{rs,mpz}. 

In order to understand the different regimes defined by the relative importance of the $\Delta$ terms in Eq.~(\ref{mixmat}), we consider four assumptions (but we contemplate the complete and general case concerning calculations about the ALP model of the GRB).

\begin{enumerate}
\item Fully polarized photons.
\item No photon absorption, namely $\lambda_{\gamma} \to \infty$.
\item Homogeneous medium.
\item Homogeneus $\bf B$ field, so that ${\bf B}(y) \equiv {\bf B}$. Thus we have the freedom to choose the $z$ axis along the direction of ${\bf B}_T$ (this fact translates 
into $\phi=0$ in Eq.~(\ref{mixmat})).
\end{enumerate}
 
With these assumptions the $\gamma \to a$ conversion probability reads
\begin{equation}
\label{convprob}
P_{\gamma \to a} ({\cal E}, y) = \left(\frac{g_{a\gamma\gamma}B_T \, l_{\rm osc} ({\cal E})}{2\pi} \right)^2 {\rm sin}^2 \left(\frac{\pi (y-y_0)}{l_{\rm osc} ({\cal E})} \right)~,
\end{equation}
where
\begin{equation}
\label{losc}     
l_{\rm osc} ({\cal E}) \equiv \frac{2 \pi}{\left[\bigl(\Delta_{zz} ({\cal E}) - \Delta_{aa} ({\cal E}) \bigr)^2 + 4 \, \Delta_{a\gamma}^2 \right]^{1/2}}~
\end{equation}
is the photon-ALP {\it oscillation length}. It is very useful to define the {\it low-energy threshold} ${\cal E}_L$ and the {\it high-energy threshold} ${\cal E}_H$ as
\begin{equation}
\label{EL}
{\cal E}_L \equiv \frac{|m_a^2 - \omega^2_{\rm pl}|}{2 g_{a \gamma \gamma} \, B_T}~,  
\end{equation}
and 
\begin{equation}
\label{EH}
{\cal E}_H \equiv g_{a \gamma \gamma} \, B_T \left[\frac{7 \alpha}{90 \pi} \left(\frac{B_T}{B_{\rm cr}} \right)^2 + \rho_{\rm CMB} \right]^{- 1}~,
\end{equation} 
respectively. For ${\cal E}_L \lesssim {\cal E} \lesssim {\cal E}_H$ the {\it strong-mixing} regime takes place and the plasma contribution, the ALP mass term, the QED one-loop effect and the photon dispersion on the CMB are negligible. In such a situation the $P_{\gamma \to a} (y)$ is maximal, energy independent and takes the form
\begin{equation}
\label{convprobSM}
P_{\gamma \to a} (y) = {\rm sin}^2 \left( \frac{g_{a\gamma\gamma} B_T}{2}  (y-y_0) \right)~.
\end{equation}
For ${\cal E} \lesssim {\cal E}_L$ the plasma contribution and/or the ALP mass term dominate and the same is true for ${\cal E} \gtrsim {\cal E}_H$ concerning the QED one-loop effect and/or the photon dispersion on the CMB: in either case we are in the {\it weak-mixing} regime and $P_{\gamma \to a} ({\cal E}, y)$ becomes energy dependent and progressively vanishes.

A final remark is important. To the extent that ALP effects are larger than those of the EBL -- which is indeed our case -- the choice of a specific EBL model is {\it totally irrelevant}, and we have taken the SL EBL model merely as an example  (see Table I above).

\section{Bounds on the ALP parameters}

The only laboratory upper bound on $g_{a \gamma \gamma}$ is provided by the CAST experiment at CERN, which is $g_{a \gamma \gamma} < 0.66 \times 10^{- 10} \, {\rm GeV}^{- 1}$ for $m_a < 0.02 \, {\rm eV}$ at the $2 \sigma$ level~\cite{cast}. Coincidentally, just the same bound has been derived from stellar evolution in globular clusters~\cite{straniero}. Other bounds from no ALP detection have been set from astrophysical observations~\cite{fermi2016,berg,conlonLim,limFabian,limJulia,limKripp,limRey2}. Type II supernovae have repeatedly been used to put bounds on the ALP parameters. In 2015 Payez {\it et al.}, using the lack of detection of reconverted photons in the Milky Way -- presumably originating from the SN1987A -- put the bound $g_{a \gamma \gamma} < 5.3 \times 10^{- 12} \, {\rm GeV}^{- 1}$ for $m_a < 4.4 \times 10^{-10} \, {\rm eV}$ (no confidence level is quoted)~\cite{payez2015}. This result has been severely criticized by Bar, Blum and D'Amico~\cite{barblumdamico2020}. In 2020 Meyer {\it et al.} published a bound~\cite{meyer2020} which was subsequently corrected as `Under the assumption that at least one SN was contained within the LAT field of view, we exclude photon-ALP couplings $g_{a \gamma \gamma} \gtrsim 2.6 \times 10^{- 11} \, {\rm GeV}^{- 1}$ for ALP masses $m_a \lesssim 3 \times 10^{- 10} \, {\rm eV}$, within a factor of $\sim 5$ within previous limits from SN1987A' (no confidence level is quoted)~\cite{meyer2020}. The strongest bounds are quite recent. One comes from the polarization measurement of the emission from magnetic white dwarfs (MWD), and reads $g_{a \gamma \gamma} < 5.4 \times 10^{- 12} \, {\rm GeV}^{- 1}$ for $m_a < 3 \times 10^{- 7} \, {\rm eV}$ at the $2 \sigma$ level~\cite{mwd}. Others arise from the analysis of blazar and/or cluster spectra in the X-ray band concerning extremely light ALPs with $m_a < {\cal O}(10^{-12}) \, {\rm eV}$ for $g_{a \gamma \gamma} < {\cal O}(10^{- 12}) \, {\rm GeV}^{- 1}$~\cite{berg,conlonLim,limFabian,limJulia,limKripp,limRey2}.  
%{\bf Even stronger bounds with $g_{a \gamma \gamma} < {\cal O}(5 \times 10^{- 13}) \, {\rm GeV}^{- 1}$ arise from the analysis of blazar and/or cluster spectra in the X-ray band concerning extremely light ALPs with $m_a < {\cal O}(10^{-12}) \, {\rm eV}$~\cite{berg,conlonLim,limFabian,limJulia,limKripp,limRey2}.}  

\section{Photon-ALP beam propagation in different regions}

Here, our aim is to investigate the photon-ALP beam propagation in each considered crossed medium: the source, the host galaxy, the extragalactic space and the Milky Way. In the following, we will skematically describe the most important properties of these media affecting the photon-ALP beam propagation, referring the reader for more details to the quoted papers which treat each specific topic.  

\subsection{Gamma-ray bursts}

We address the case in which multi-TeV photons are produced in the external forward shock driven by the GRB jet into the surrounding medium (afterglow emission), in analogy with all the other GRBs detected at $\sim$TeV energies. Before leaving the source, photons must cross the downstream region, which has a comoving length given by $\Delta R^{\prime}\sim R/\Gamma$, where $R$ is the distance from the central engine and $\Gamma$ is the bulk Lorentz factor. For the surrounding medium we consider both an ISM-like constant number density $n(R)=n_0$ and a wind-like density $n(R) = A R^{-2}$ shaped by the wind of the progenitor star. In the latter case, $A$ is related to the mass loss rate of the progenitor's star $\dot M$ and to the velocity of the wind $v_w$ by $A =\dot M/( 4\,\pi\,m_p\,v_w)$ where $m_p$ is the proton mass. We normalize the value of $A$ to a mass loss rate of $10^{-5}$ solar masses per year and a wind velocity of $10^3$\,km\,s$^{-1}$. Accordingly we get  $A=3\times10^{35}\,A_\star$\,cm$^{-1}$. Therefore, we finally obtain $n(R) = 3\times10^{35}A_\star R^{-2} \, \rm cm^{-1}$. In the comoving frame (namely the frame at rest with the shocked fluid) the density is given by $n^{\prime} \sim n \,\Gamma$ and the magnetic field strength is $B^{\prime}=\sqrt{32\pi\epsilon_B m_p c^2 n(R)}\Gamma$, where $\epsilon_B$ is the fraction of shock-dissipated energy gained by the magnetic field, and $c$ the speed of light.

Since we are interested in understanding whether or not the conversion probability in the emitting region can be important, we adopt parameters that maximize the photon survival probability, and verify that even in this optimistic case the conversion within the emitting region can be neglected. 

TeV emission was detected in the first 2000\,s from the initial burst, but the exact time when the 18\,TeV photon was detected is unknown at the time of writing. Since an important term entering the calculations is $B^{\prime}\Delta R\sim B^{\prime} R/\Gamma$, we choose a time that maximizes its value within the first 2000\,seconds. As we will show later, this term is maximized at 2000\,seconds, when the blast-wave is in the deceleration phase. We then adopt for the bulk Lorentz factor the equation provided in \cite{bm76} and valid during the deceleration phase. Its value at a given time depends only on the kinetic energy of the jet ${\cal E}_k$ and on the density $n(R)$ of the external unshocked medium. For simplicity, the relation between time in the observer frame and radius where the emission is produced is approximated using the equation $R \sim 4 c \, \Gamma^2 (1+z)$, since more detailed equations (see e.g. \cite{navasironi,derishevpiran,derishev}) do not change our results. Denoting by $t_{\rm obs}$ the time in the observer frame, we find that $B^{\prime} R/\Gamma$ is proportional to ${\cal E}_k^{1/4} \, n_0^{1/4} \, \epsilon_B^{1/2} \,t_{\rm obs}^{1/4}$ in a constant medium, and to $\epsilon_B^{1/2} A_\star^{1/2}$ in a wind medium (note that in the latter case it is constant in time). We then assume large -- but still reasonable -- values for the jet energy, the medium density and the fraction $\epsilon_B$ of shock-dissipated energy gained by the magnetic field, and consider the value of $B^{\prime} \, R/\Gamma$ at $t_{\rm obs}=2000$\,s. In particular, calculations provided in the Letter are given for an homogeneous medium and ${\cal E}_k=10^{54}$ \, {\rm erg}, $n_0=10$\,cm$^{-3}$, $A_\star=10$ and $\epsilon_B=10^{-3}$. We find that the value of 
$B^{\prime} R/\Gamma$ is much larger for the case of constant density and is maximal at 2000 seconds, when $B^{\prime} \sim 2$\,G, $R \sim 2 \times10^{17}$\,cm, $\Gamma=45$, and $n^{\prime} \sim 450$\,cm$^{-3}$. 

By using these values, we compute the transfer matrix for the photon-ALP beam propagation in the source region ${\cal U}_1 ({\cal E}; y_2, y_1)$, where $y_2$ and $y_1$ denote the position of the border of the GRB and of the production region, respectively. Because of the above-discussed values assumed by $R$ and $\Gamma$, the relevant propagation length of photon-ALP beam in the source is $\Delta R^{\prime} \sim R/\Gamma \sim 5 \times 10^{15} \, \rm cm$. Following a similar calculation scheme as the one reported in~\cite{gtre2019}, we have found that $\Delta R^{\prime}$ turns out to be too small to produce an efficient photon-ALP conversion inside the source. As a result, we obtain ${\cal U}_1 ({\cal E}; y_2, y_1) \simeq 1$, where $y_2$ and $y_1$ denote the position of the border of the GRB and of the production region, respectively.

\subsection{Host galaxy}

Although no detailed information about the nature of the galaxy hosting GRB 221009A is available in the literature, currently we know that it should be a disc-like galaxy seen edge-on -- so that the photon/ALP beam travels inside the disk -- with the GRB located close to the nuclear region~\cite{GRB221009Ahost}. The latter fact is in agreement with the observations that long GRBs are normally hosted in the central region of their host galaxy, as suggested by~\cite{GRBposition,GRBposition2}. Accordingly,  we place GRB 221009A in the neighborhood of the galactic center. 

The two most obvious possibilities about the nature of the host are represented by spirals and starburst galaxies. For both a spiral and a starburst galaxy hosting the GRB we assume a luminous radius $y_3 = 20 \, \rm kpc$. Moreover, rather conservatively with respect to an efficient photon-ALP mixing, we assume an electron number density $n_{{\rm host}, \, e}=1 \, \rm cm^{-3}$ for spirals~\cite{SpiralBrev} and $n_{{\rm host}, \, e} = 1000 \, \rm cm^{-3}$ for starbursts~\cite{LopezRodriguez2021}.

The overall magnetic field ${\bf B}_{\rm host}$ consists of three contributions. 
\begin{enumerate}
\item The regular component ${\bf B}_{\rm reg}$.
\item The large-scale anisotropic turbulent component ${\bf B}_{\rm aniso}$.
\item The small-scale isotropic turbulent component ${\bf B}_{\rm iso}$.
\end{enumerate}
Below, we shall discuss them separately (for a review see~\cite{SpiralBrev}).

Concerning the behavior along the $y$ direction -- which is a radial direction with respect to the host -- of any magnetic component $\bf B$, we assume an exponential profile~\cite{Heesen2023} and we model the magnetic field coherence by employing a Kolmogorov-type turbulence power spectrum $M(k)\propto k^q$, with $k$ the wave number in the range $[k_L,k_H]$ and index $q=-11/3$. The lower and upper limits of the latter interval $k_L = 2\pi/\Lambda_{\rm max}$ and $k_H = 2\pi/\Lambda_{\rm min}$ represent the minimal and maximal turbulence scales, respectively. As a result, by taking the appropriate  values of the parameters concerning strength and coherence, each generic component ${\bf B}$ of ${\bf B}_{\rm host}$ can be modeled by
\begin{equation}     
\label{Bprofile}
B(y) = {\cal B} \left( B_0,k,q,y \right) \exp{(-y/r_0)}~,
\end{equation}
where ${\cal B}$ represents the spectral function accounting for the Kolmogorov-type turbulence of the host magnetic field (see e.g.~\cite{meyerKolm} for more details), while $B_0$ and $r_0$ represent the central magnetic field strength and the radial scale length, respectively, of every magnetic field component. 

\medskip

\noindent {\bf Spiral galaxies:} In these galaxies ${\bf B}_{\rm reg}$ gives the most relevant contribution to the magnetic field, as far as photon-ALP interaction is concerned. Spirals are characterized by a lower star formation and consequently smaller supernova explosion rate -- as compared to starburst galaxies -- which is the most important source for the other two magnetic components. As a result, the turbulent components of ${\bf B}_{\rm host}$ -- although present -- contribute less to the photon-ALP interaction with respect to ${\bf B}_{\rm reg}$: therefore, in order to be conservative we take only ${\bf B}_{\rm reg}$ into account. Often, a coherence length of $L_{\rm dom}^{\rm host} \sim 10 \, \rm kpc$ with $B_{\rm host} \sim 5 \, \mu{\rm G}$ is assumed for spirals in the literature~\cite{Fletcher2010}. However, we model ${\bf B}_{\rm host}$ more accurately by varying $L_{\rm dom}^{\rm host}$ within a more physical smaller range (see e.g.~\cite{SpiralBrev}) by assuming a Kolmogorov type turbulence power spectrum. In particular, in order to describe the coherence structure, we accordingly take $\Lambda_{\rm min}= 2 \, \rm kpc$ and $\Lambda_{\rm max}= 4 \, \rm kpc$, and concerning the radial profile of ${\bf B}_{\rm reg}$ we employ Eq.~(\ref{Bprofile}) with $B \to B_{\rm reg} \equiv  B_{\rm host}$, $B_0 \to B_{{\rm reg}, \, 0}= 20 \, \mu{\rm G}$ and, conservatively $r_0 \to r_{{\rm reg}, \, 0}= 15 \, \rm kpc$. Specifically, $r_{{\rm reg}, \, 0}$ can be $\sim 6 \, \rm kpc$ in the halo and up to $\sim 20 \, \rm kpc$ when the host is seen edge-on~\cite{SpiralBrev}, which is indeed our case. The value chosen for $B_{{\rm reg}, \, 0}$ is in agreement with that of typical spirals, and the one for $r_{{\rm reg}, \, 0}$ is consistent with a galaxy observed close to the edge-on direction (see e.g.~\cite{SpiralBrev,Heesen2023}). In Fig.~\ref{Bspiral}, we report a particular realization of the component along the $x$-axis of ${\bf B}_{\rm host}$ for a spiral galaxy with the above properties.
\begin{figure}
\begin{center}
\includegraphics[width=.50\textwidth]{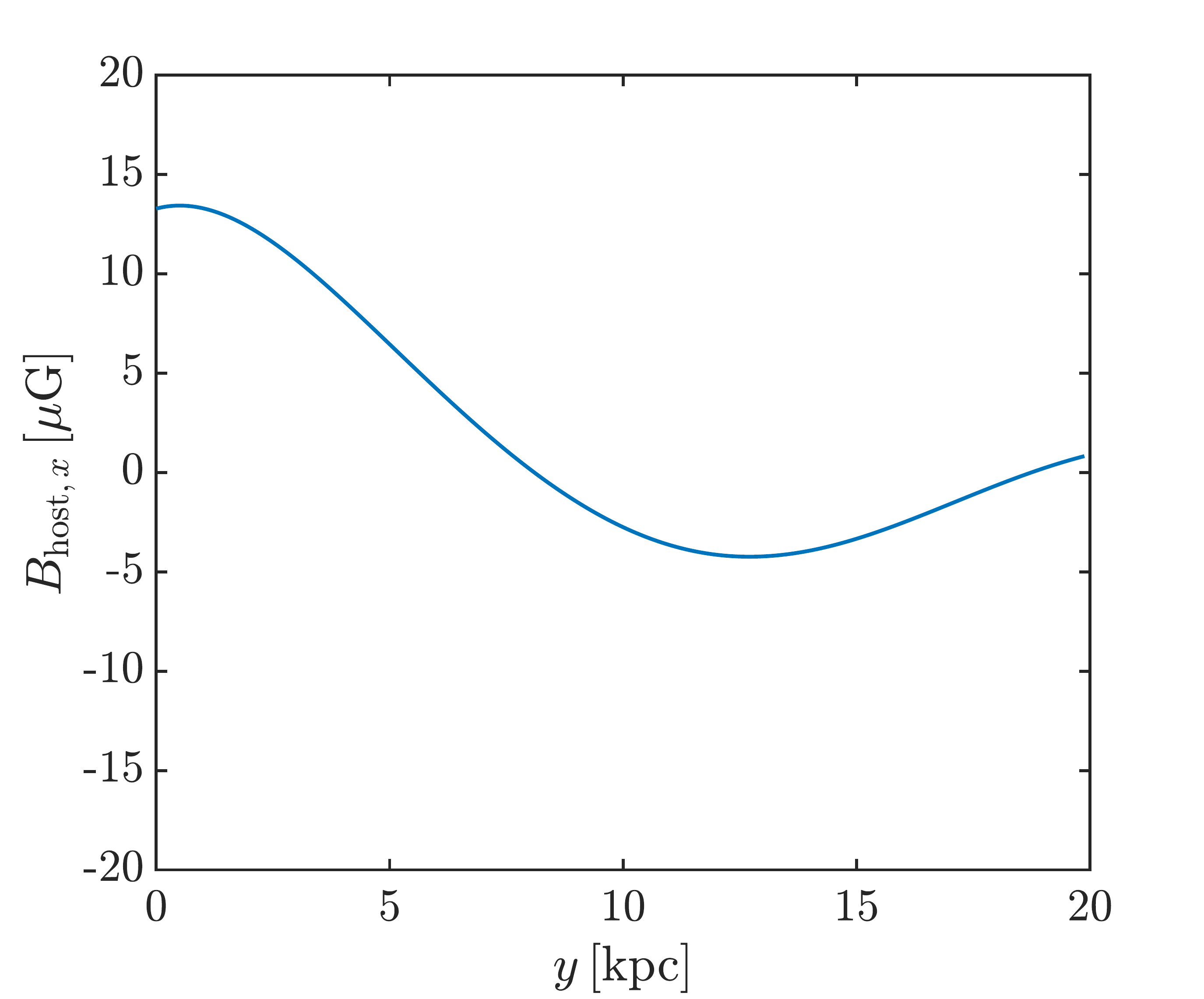}
\end{center}
\caption{\label{Bspiral} Component along the $x$-axis of a realization of the spiral galaxy magnetic field $B_{{\rm host}, \, x}$ with respect to the galactic distance $y$.}
\end{figure}

\medskip

\noindent {\bf Starburst galaxies:} As far as these galaxies are concerned, all the 
above-defined magnetic field components are in principal relevant. The choice of the benchmark parameters is inspired by M82~\cite{LopezRodriguez2021}, which is considered the prototype and also an example of an average starburst galaxy~\cite{Thompson2006}. In analogy with spirals we employ a Kolmogorov spectrum to describe the coherence of ${\bf B}_{\rm reg}$ and we assume $\Lambda_{\rm min} = 1.5 \, \rm kpc$ and $\Lambda_{\rm max}= 3 \, \rm kpc$, which are values lower than those taken for the spiral galaxies, in order to be more conservative. As in the case of spirals, we assume to observe the starburst galaxy along the edge-on direction, again in agreement with the case of GRB 221009A. Therefore, we describe the radial profile of ${\bf B}_{\rm reg}$ by using Eq.~(\ref{Bprofile}) with $B \to B_{\rm reg}$, $B_0 \to B_{{\rm reg}, \, 0}= 50 \, \mu{\rm G}$ and, conservatively, $r_0 \to r_{{\rm reg}, \, 0}= 10 \, \rm kpc$ (see e.g.~\cite{SpiralBrev,Heesen2023} for the parameter choice). Concerning ${\bf B}_{\rm aniso}$ and ${\bf B}_{\rm iso}$, it is important to define the average turbulent coherence length $\delta \sim 70 \, \rm pc$ driven by supernova explosions, which take place at scales of $(50-100) \, \rm pc$ in nearby galaxies (see e.g.~\cite{Elmegreen2004,Haverkorn2008}). In order to model ${\bf B}_{\rm aniso}$, we describe its coherence by assuming a Kolmogorov type turbulence power spectrum in a similar way as considered above but with $\Lambda_{\rm min}= 70 \, \rm pc$ and $\Lambda_{\rm max}= 150 \, \rm pc$, since ${\bf B}_{\rm aniso}$ is believed to be generated by the galactic winds on scales equal or larger than $\delta$~\cite{LopezRodriguez2021}. Even for the radial profile of ${\bf B}_{\rm aniso}$ we employ Eq.~(\ref{Bprofile}) with $B \to B_{\rm aniso}$, $B_0 \to B_{{\rm aniso}, \, 0}= 700 \, \mu{\rm G}$ and $r_0 \to r_{{\rm aniso}, \, 0}= 1 \, \rm kpc$ (see e.g.~\cite{LopezRodriguez2021} for the parameter choice). Instead, the typical coherence length of ${\bf B}_{\rm iso}$ is smaller than $\delta$~\cite{LopezRodriguez2021}. Therefore, the contribution of ${\bf B}_{\rm iso}$ is much less important for the photon-ALP interaction with respect to that of ${\bf B}_{\rm aniso}$: for this reason and in order to be conservative, we do not take the contribution of ${\bf B}_{\rm iso}$ into account. As a consequence, the total magnetic field ${\bf B}_{\rm host}$ relevant for the photon-ALP conversion in the starburst galaxy is represented by the sum of ${\bf B}_{\rm reg}$ and of ${\bf B}_{\rm aniso}$. A particular realization of the component along the $x$-axis of ${\bf B}_{\rm host}$ in a starburst galaxy with the above-described properties is reported in Fig.~\ref{Bstarburst}, from which we see that the turbulent nature of ${\bf B}_{\rm aniso}$ dominates for galactic distances smaller than $\sim 3 \, \rm kpc$, while the influence of ${\bf B}_{\rm reg}$ is prominent on larger scales.

\begin{figure}
\begin{center}
\includegraphics[width=.50\textwidth]{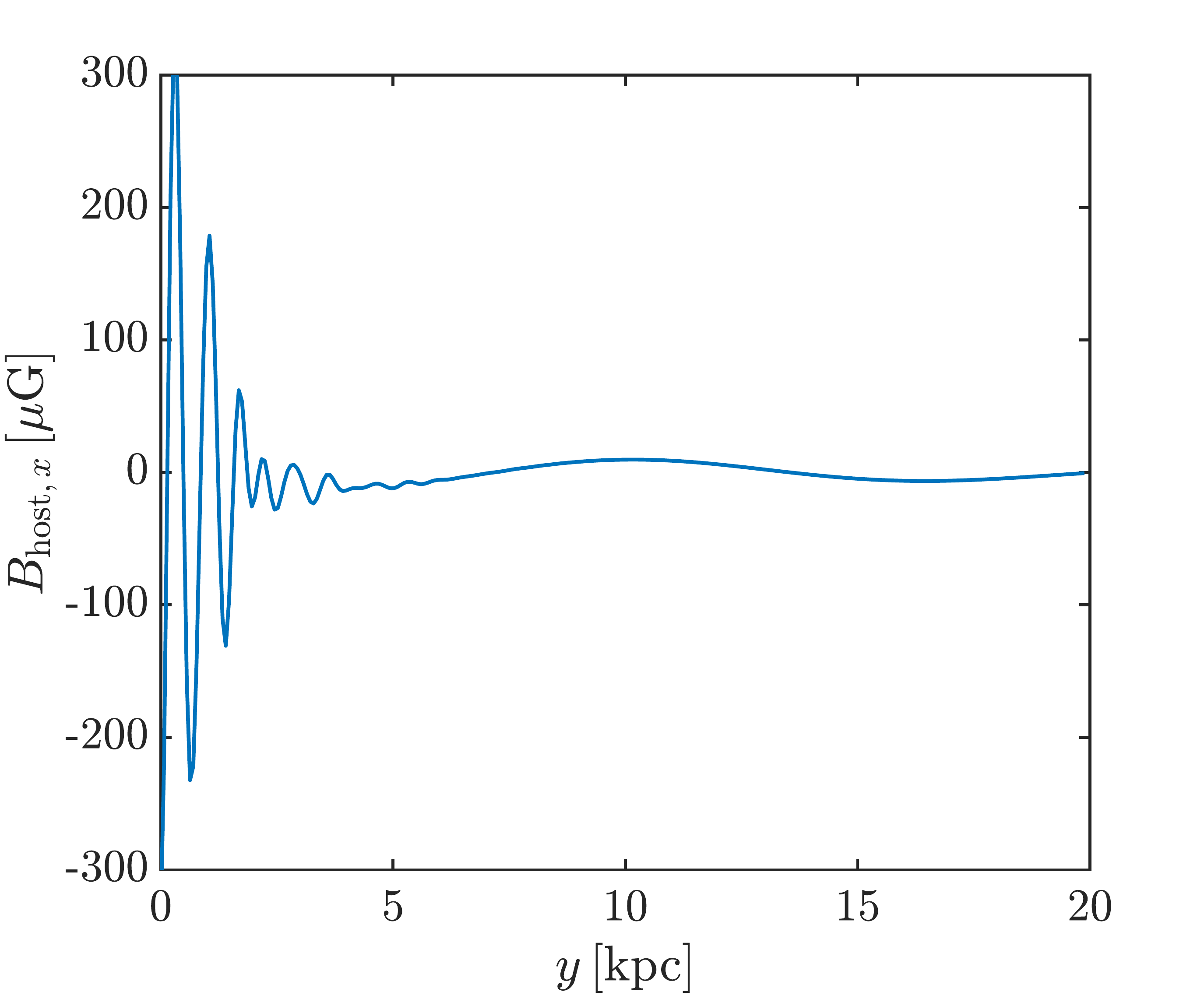}
\end{center}
\caption{\label{Bstarburst} Component along the $x$-axis of a realization of the starburst galaxy magnetic field $B_{{\rm host}, \, x}$ with respect to the galactic distance $y$.}
\end{figure}

Following the same strategy developed in~\cite{galantiPol} concerning photon-ALP conversion inside galaxy clusters, we can compute the transfer matrix in the host galaxy ${\cal U}_2 ({\cal E}; y_3, y_2)$ in the case of either a spiral or a starburst, where $y_3$ denotes the position of the external luminous galaxy radius.

%\newpage

\subsection{Extragalactic space}

Unfortunately, our knowledge of the extragalactic magnetic field ${\bf B}_{\rm ext}$ is very poor. All we know is that $B_{\rm ext}$ has to lie in the range $10^{- 7} \, {\rm nG} \lesssim B_{\rm ext} \lesssim 1.7 \, {\rm nG}$ on the scale of ${\cal O} (1) \, {\rm Mpc}$~\cite{neronovvovk,durrerneronov,upbbext}. 

Nevertheless, according to the current wisdom ${\bf B}_{\rm ext}$ can be modeled as a domain-like network, wherein ${\bf B}_{\rm ext}$ is assumed to be homogeneous over a whole domain of size $L_{\rm dom}^{\rm ext} $ equal to its coherence length, with ${\bf B}_{\rm ext}$ changing randomly its direction from one domain to the next, keeping approximately the same strength~\cite{kronberg1994,grassorubinstein2001}. Accordingly, the beam propagation becomes a {\it random process}, and only a single realization at once can be observed. Moreover, it is generally assumed that such a change of direction is abrupt, because then the beam propagation equation is easy to solve. The physics behind this scenario -- called {\it domain-like sharp-edges} (DLSHE) -- relies upon outflows from primeval galaxies, further amplified by turbulence~\cite{reessetti1968,hoyle1969,kronbergleschhopp1999,furlanettoloeb2001}. Common benchmark values are $B_{\rm ext} = {\cal O} (1) \, {\rm nG}$ on a coherence length ${\cal O} (1) \, {\rm Mpc}$, whereas $L_{\rm dom}^{\rm ext} = {\cal O} (1) \, {\rm Mpc}$ (for more details, see~\cite{galantironcadelli20118prd}). In order to be definite, we choose $B_{\rm ext} = 1 \, {\rm nG}$ and $L_{\rm dom}^{\rm ext}$ in the range $(0.2-10) \, \rm Mpc$ and with $\langle L_{\rm dom}^{\rm ext}  \rangle = 2 \, \rm Mpc$. 

However, the abrupt change in direction at the interface of two adjacent domains leads to a failure of the DLSHE model at the energies considered here, owing to photon dispersion on the CMB~\cite{raffelteffect}. A way out of this difficulty is to smooth out the sharp edges of the domains, in such a way that the components of ${\bf B}_{\rm ext}$ change continuously across the interface, thereby leading to the {\it domain-like smooth-edges} (DLSME) model, built up in~\cite{galantironcadelli20118prd,kartavtsev}. 

Coming back to the propagation of a photon-ALP beam in extragalactic space within the DLSME model, $\gamma \leftrightarrow a$ oscillations imply that the photons behave in two-fold fashion. When they are true photons, they undergo EBL absorption, but when they are ALPs they do not. So, the effective optical depth $\tau_{\rm ALP} ({\cal E})$ gets reduced. The crux of the argument is that the photon survival probability now becomes
\begin{equation}
P_{\rm ALP} ({\cal E}; \gamma \to \gamma) = e^{- \, \tau_{\rm ALP}(\cal E)}~,  
\label{a3}
\end{equation}
and so even a small decrease of $\tau_{\rm ALP} ({\cal E}) $ gives rise to a large increases of $P_{\rm ALP} ({\cal E}; \gamma \to \gamma)$~\cite{drm2007}. Moreover, photon dispersion on the CMB induces small energy oscillations in the beam. This analysis has been done in~\cite{grjhea}, and yields ${\cal U}_3 ({\cal E}; y_4, y_3)$, where $y_4$ denotes the position of the outer luminous edge of the Milky Way. 

Thus -- given the uncertainty about $B_{\rm ext}$ -- we shall consider also the alternatively and extremely conservative case of $B_{\rm ext} < 10^{- 15} \, {\rm G}$. 

\subsection{Milky Way}

Clearly, the beam propagation in the Milky Way -- as triggered by $\gamma \leftrightarrow a$ oscillations -- depends both on the morphology of the magnetic field ${\bf B}_{\rm MW}$ and on the electron number density $n_{{\rm MW}, \, e}$, which fixes in turn the plasma frequency $\omega_{{\rm MW, \, pl}}$. 

We believe that the best model for ${\bf B}_{\rm MW}$ is the one developed by Jansson and Farrar~\cite{jansonfarrar1,jansonfarrar2,BMWturb}. It includes a disc and a halo -- in both cases ${\bf B}_{\rm MW}$ is parallel to the Galactic plane -- and a poloidal ‘X-shaped’ component at the Galactic centre. Even though this model contains a regular and a turbulent components, only the former is relevant for the present needs~\cite{remark2}. We have tested the robustness of our results by considering also the alternative model of Pshirkov {\it et al.}~\cite{pshirkov2011}. Even with some little modifications our findings are basically unchanged.

As far as $n_{{\rm MW}, \, e}$ is concerned, its overall best estimate is $n_{{\rm MW}, \, e} \simeq 1.1 \times 10^{- 2} \, {\rm cm}^{- 3}$~\cite{yaomanchesterwang 2017}, which results in $\omega_{{\rm MW, \, pl}} \simeq 3.9 \times 10^{- 12} \, {\rm eV}$. Correspondingly, ${\cal U}_4 ({\cal E}; y_5, y_4)$ -- where $y_5$ is the position of the Earth -- can be evaluated as in Section 3.4 of~\cite{gtre2019}.

\subsection{Total photon-ALP beam propagation}

As explained in the text -- and according to quantum mechanics -- the total transfer matrix is 
\begin{equation}
{\cal U} ({\cal E}; y_5, y_1) = \prod_{i = 1}^4 {\cal U}_i ({\cal E}; y_{i + 1}, y_i)
\label{14022023a}
\end{equation}
and the total survival probability is given by 
\begin{equation}
P_{\rm ALP} ({\cal E}; \gamma \to \gamma) = \sum_{i = x, z} {\rm Tr} \bigl[\rho_i \, {\cal U} ({\cal E}; y_5, y_1) \rho_{\rm unp} \, {\cal U}^{\dagger} ({\cal E}; y_5, y_1) \bigr]~,
\label{14022023b}
\end{equation}
with
\begin{equation}
\label{densphot}
{\rho}_x = \left(
\begin{array}{ccc}
1 & 0 & 0 \\
0 & 0 & 0 \\
0 & 0 & 0 \\
\end{array}
\right)~, \,\,\,\,\,\,\,\,
{\rho}_z = \left(
\begin{array}{ccc}
0 & 0 & 0 \\
0 & 1 & 0 \\
0 & 0 & 0 \\
\end{array}
\right)~,
\end{equation}
accounting for pure photon states with polarization along the $x$ and $z$ directions, respectively, and
\begin{equation}
\label{densunpol}
{\rho}_{\rm unp} = \frac{1}{2} \left(
\begin{array}{ccc}
1 & 0 & 0 \\
0 & 1 & 0 \\
0 & 0 & 0 \\
\end{array}
\right)~,
\end{equation}
describing initially unpolarized photons.

\newpage

\section{Extended results}

\subsection{Spiral galaxy}

\begin{figure*}[h]
\begin{center}
\includegraphics[width=.48\textwidth]{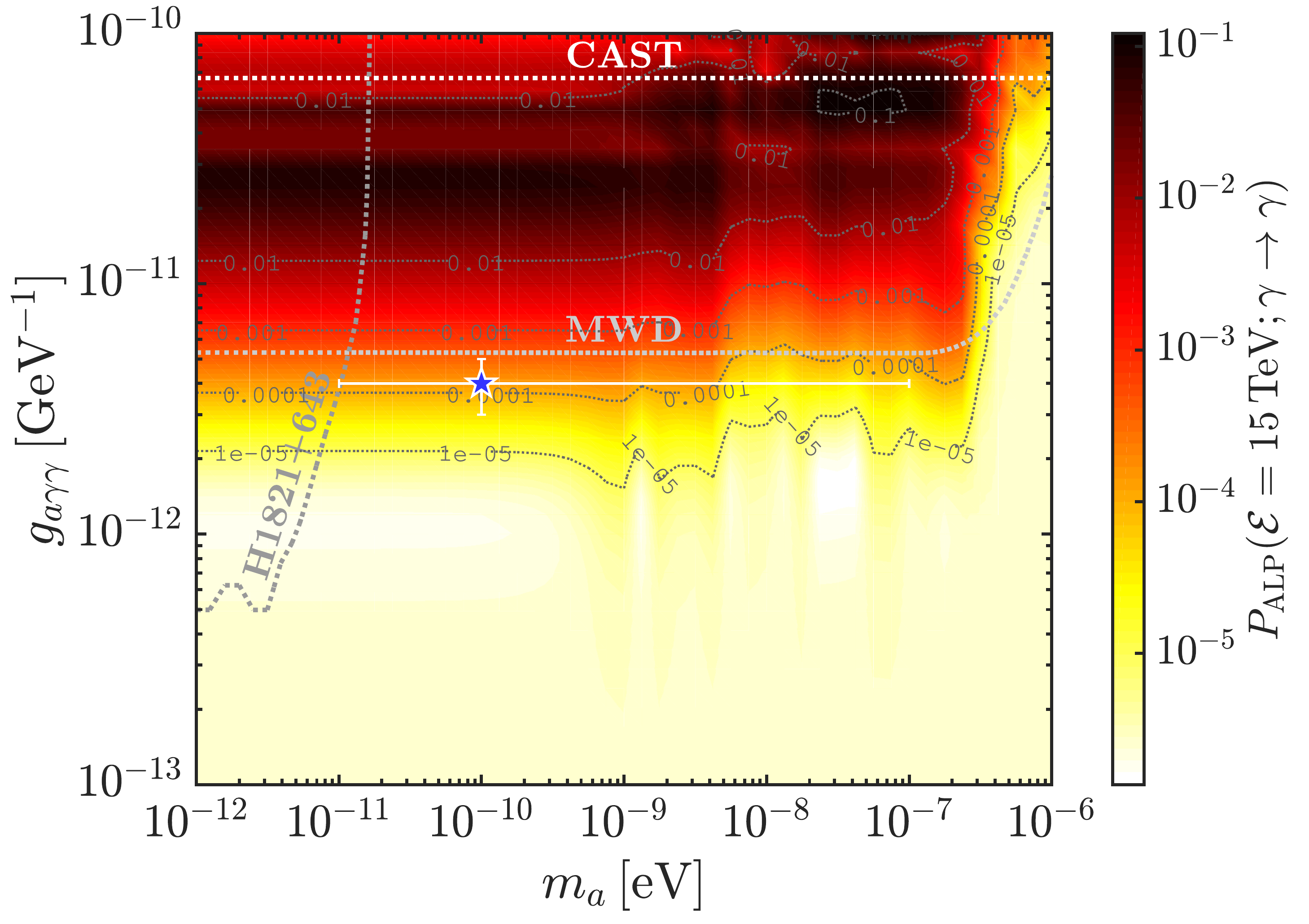} \hspace{3mm} \includegraphics[width=.48\textwidth]{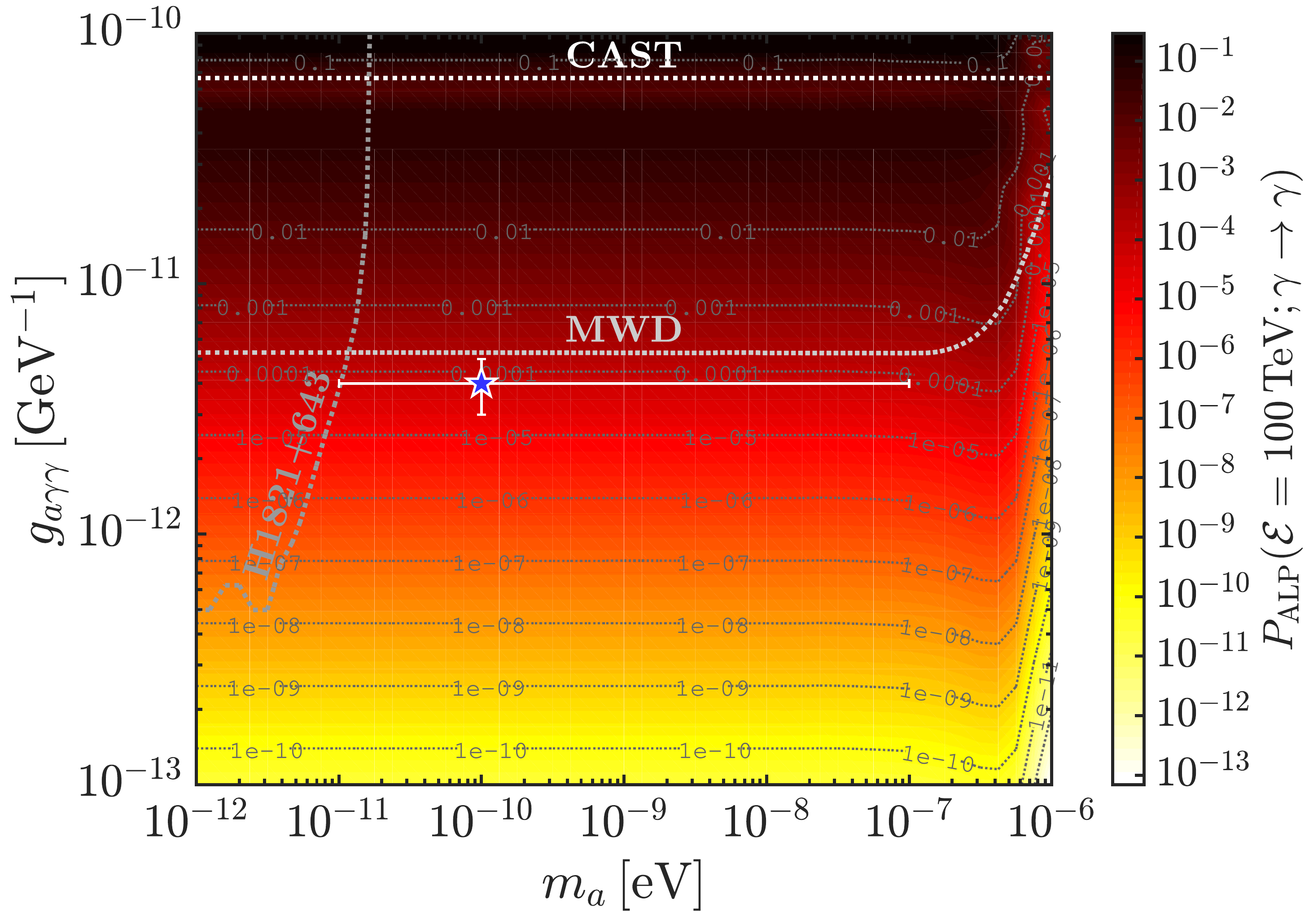}
\includegraphics[width=.48\textwidth]{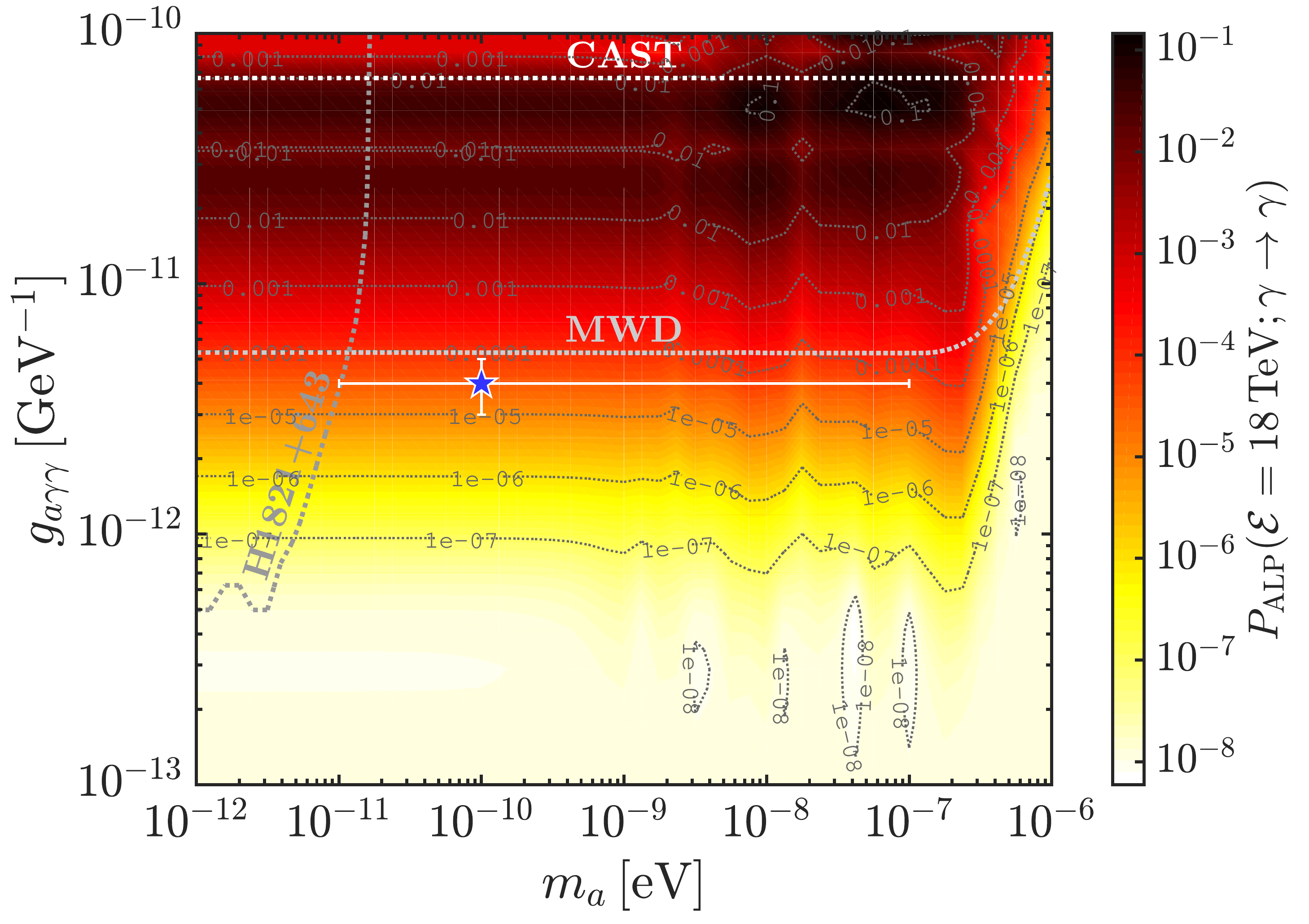} \hspace{3mm} \includegraphics[width=.48\textwidth]{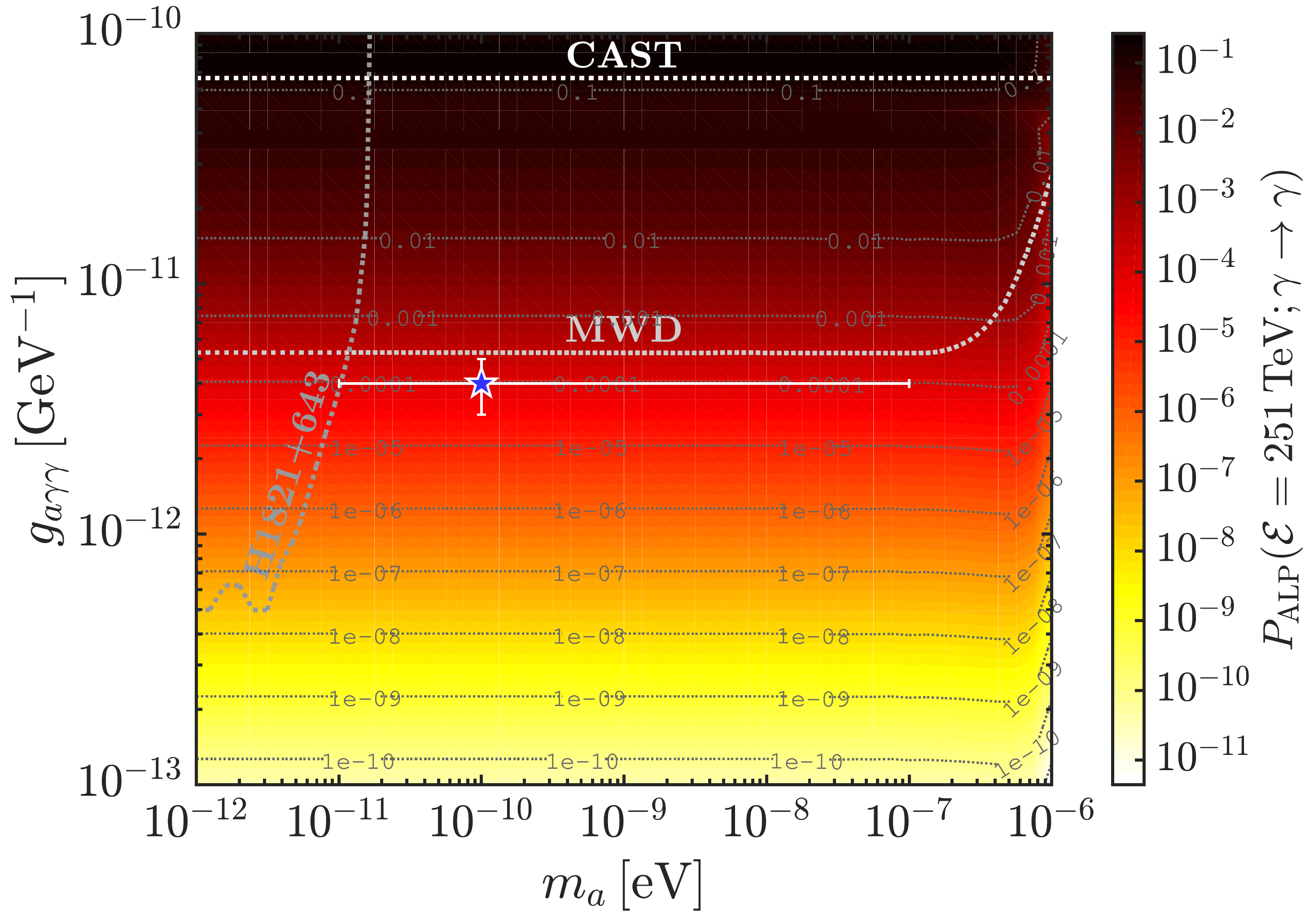}
\end{center}
\caption{\label{parSpaceSpiralBig} Photon survival probability $P_{\rm ALP}({\cal E}; \gamma \to \gamma)$ at energies ${\cal E} = 15 \, \rm TeV$ (upper-left panel), ${\cal E} = 18 \, \rm TeV$ (lower-left panel), ${\cal E} = 100 \, \rm TeV$ (upper-right panel) and ${\cal E} = 251 \, \rm TeV$ (lower-right panel), as a function of the ALP mass $m_a$ and of the photon-ALP coupling $g_{a\gamma\gamma}$ by assuming the SL EBL model, $B_{\rm ext}=1 \, \rm nG$ and a spiral hosting galaxy. The blue star with error bar represents our choice for the ALP parameters: $m_a=10^{-10} \, \rm eV$ and $g_{a\gamma\gamma}=4 \times 10^{-12} \, \rm GeV^{-1}$. The CAST bound~\cite{cast}, that coming from magnetic white dwarf (MWD) polarization~\cite{mwd} and the one derived from H1821+643~\cite{limJulia} are also plotted.}
\end{figure*}      

\begin{figure}[H]
\begin{center}
\includegraphics[width=.45\textwidth]{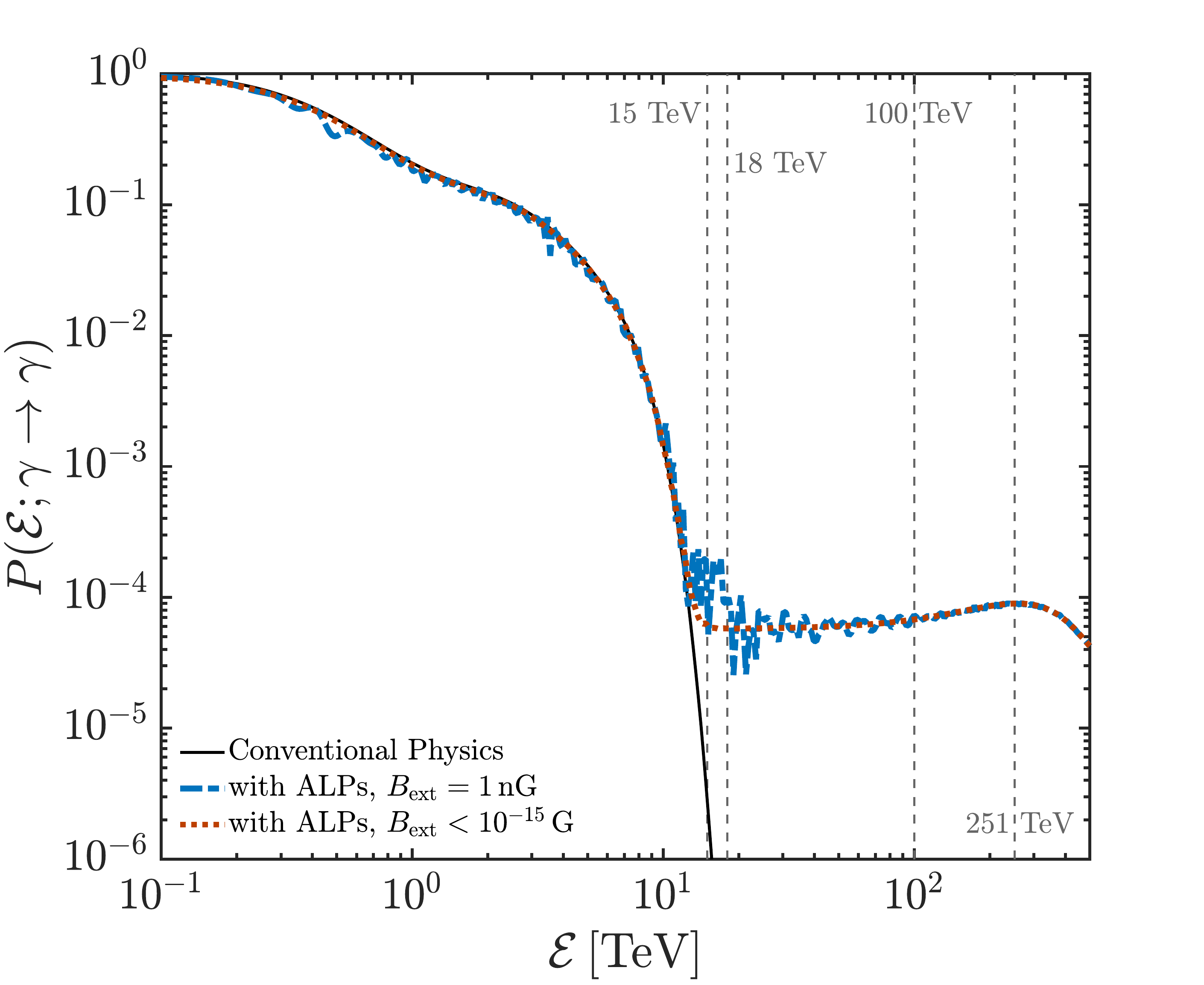}
\end{center}
\caption{\label{survProbFigSpiral} Photon survival probability $P ({\cal E}; \gamma \to \gamma)$ versus energy ${\cal E}$ in conventional physics and when ALPs (for $B_{\rm ext} = 1 \, \rm nG$ and $B_{\rm ext} < 10^{-15} \, \rm G$) are considered for the case of a spiral hosting galaxy and employing the SL EBL model. We assume $m_a=10^{-10} \, \rm eV$ and $g_{a\gamma\gamma}=4 \times 10^{-12} \, \rm GeV^{-1}$.}
\end{figure} 

\begin{figure}[H]
\begin{center}
\includegraphics[width=.5\textwidth]{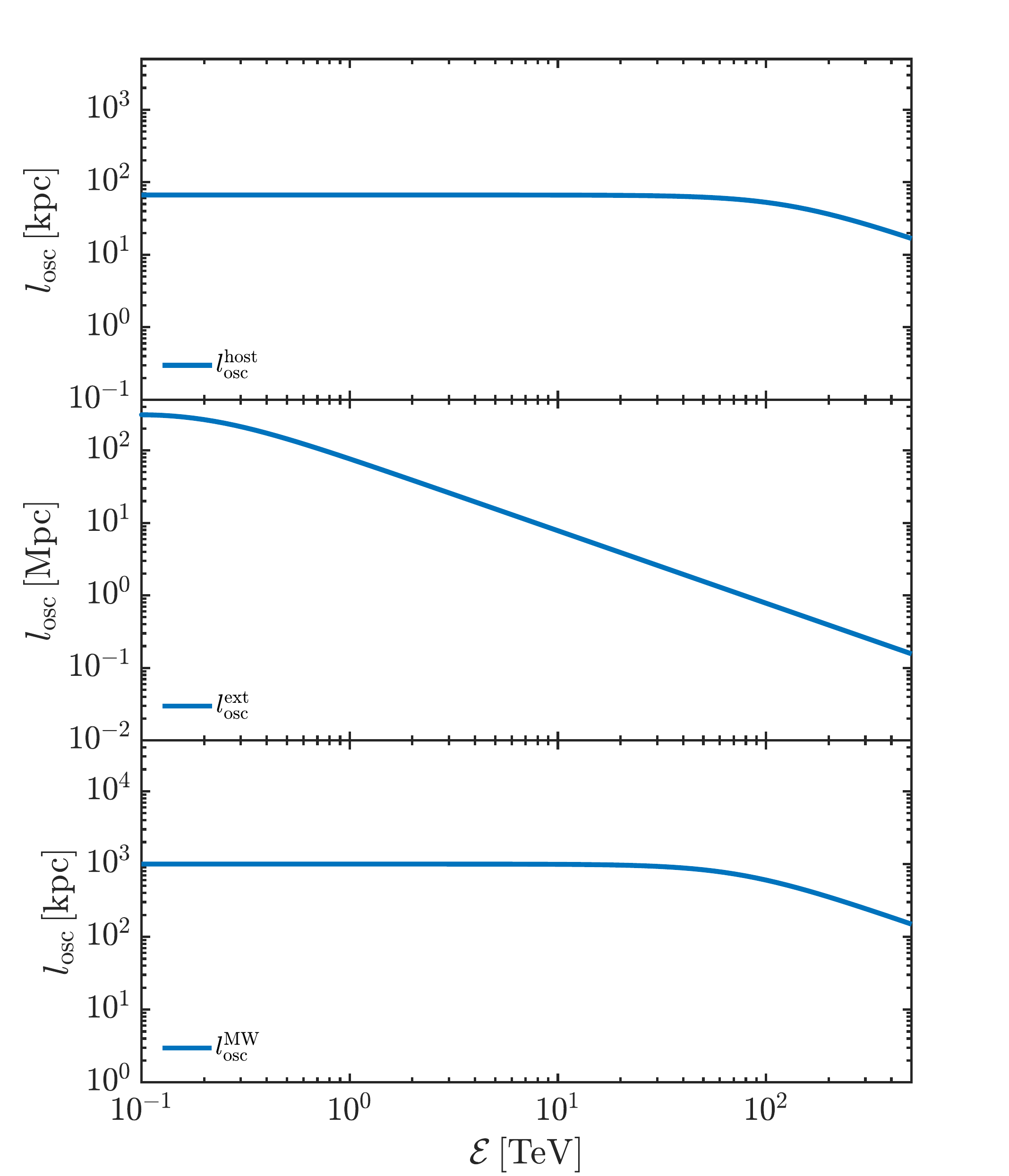}
\end{center}
\caption{\label{LoscSpiralPlot} Oscillation length $l_{\rm osc}$ versus energy ${\cal E}$ in the different crossed regions: the spiral galaxy hosting the GRB ($l_{\rm osc}^{\rm host}$, upper panel), the extragalactic space ($l_{\rm osc}^{\rm ext}$ with $B_{\rm ext} = 1 \, \rm nG$, central panel) and the Milky Way ($l_{\rm osc}^{\rm MW}$, lower panel). We assume $m_a=10^{-10} \, \rm eV$ and $g_{a\gamma\gamma}=4 \times 10^{-12} \, \rm GeV^{-1}$.}
\end{figure} 

\begin{figure}[H]
\begin{center}
\includegraphics[width=.5\textwidth]{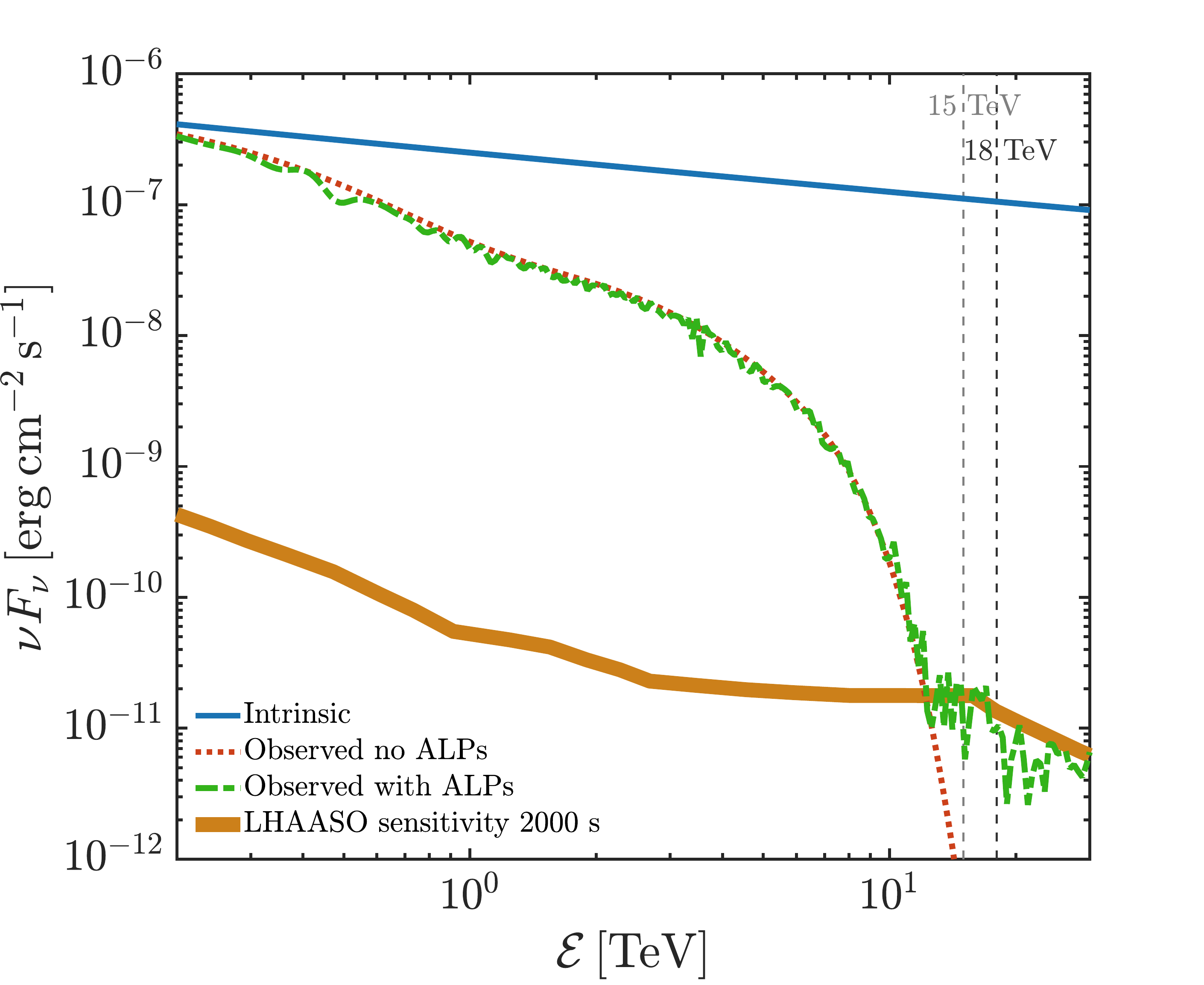}
\end{center}
\caption{\label{SpectrumSpiralBig} Intrinsic average spectrum of GRB 221009A as measured by LHAASO~\cite{grb221009aSpectrum} extended up to $\sim 20 \, \rm TeV$ and corresponding observed one versus energy ${\cal E}$ within conventional physics and when ALP effects are taken into account by assuming the SL EBL model, $B_{\rm ext}=1 \, \rm nG$, a spiral hosting galaxy, $m_a=10^{-10} \, \rm eV$ and $g_{a\gamma\gamma}=4 \times 10^{-12} \, \rm GeV^{-1}$. The LHAASO sensitivity at $2000 \, \rm s$ is also shown.}
\end{figure} 

\newpage

\subsection{Starburst galaxy}

\begin{figure*}[h]
\begin{center}
\includegraphics[width=.48\textwidth]{paramSpaceStarburst15TeV.pdf} \hspace{3mm} \includegraphics[width=.48\textwidth]{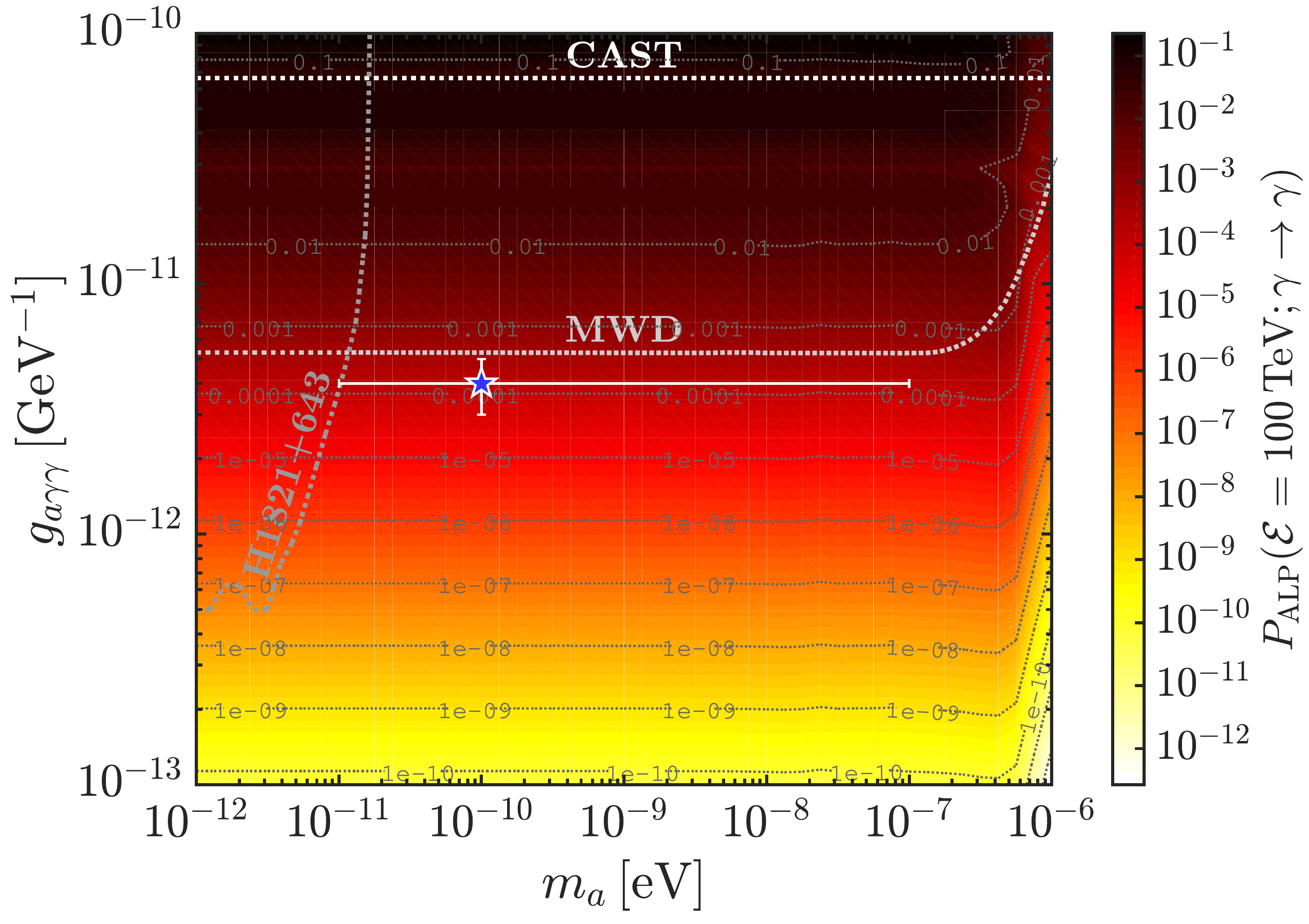}
\includegraphics[width=.48\textwidth]{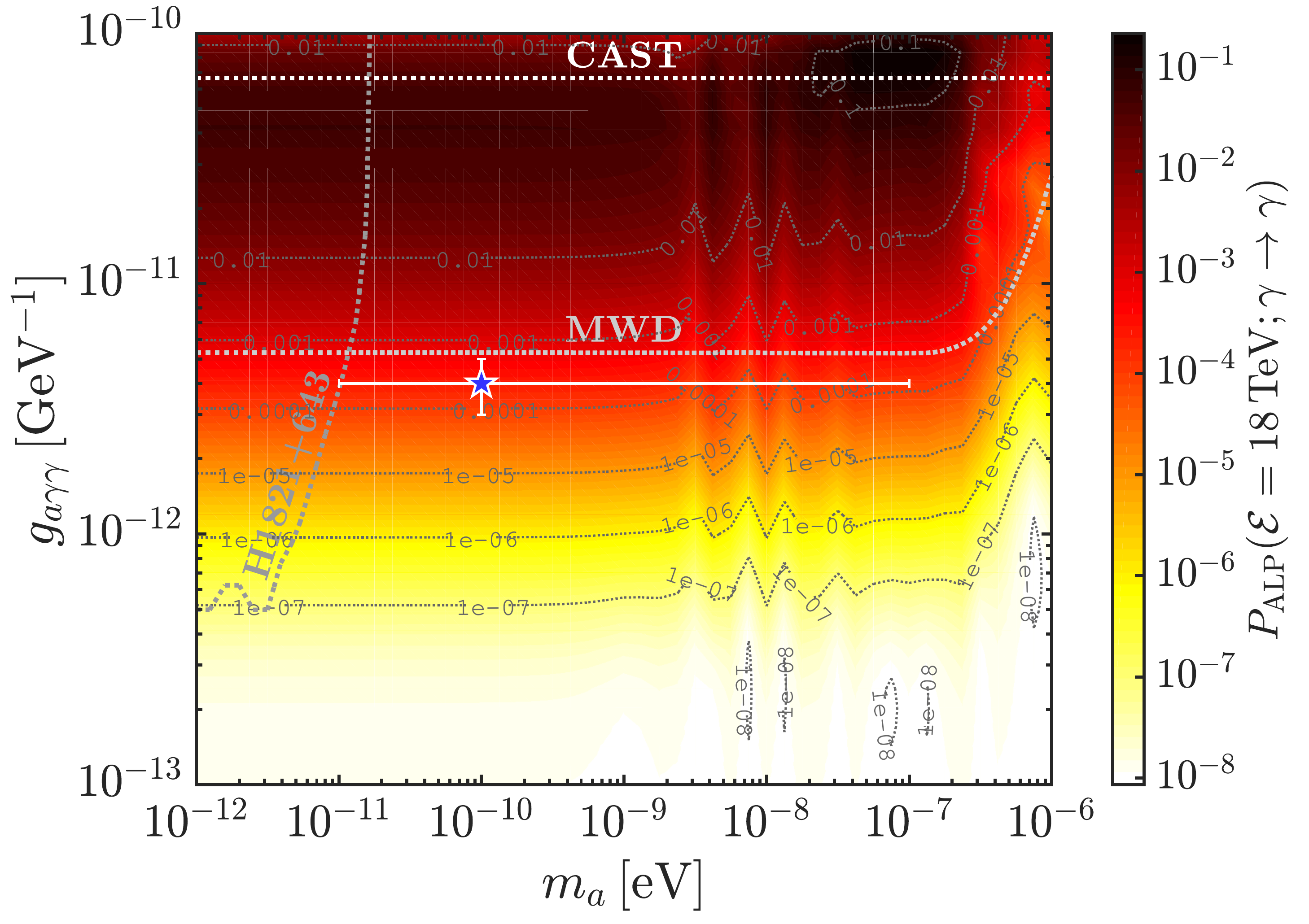} \hspace{3mm} \includegraphics[width=.48\textwidth]{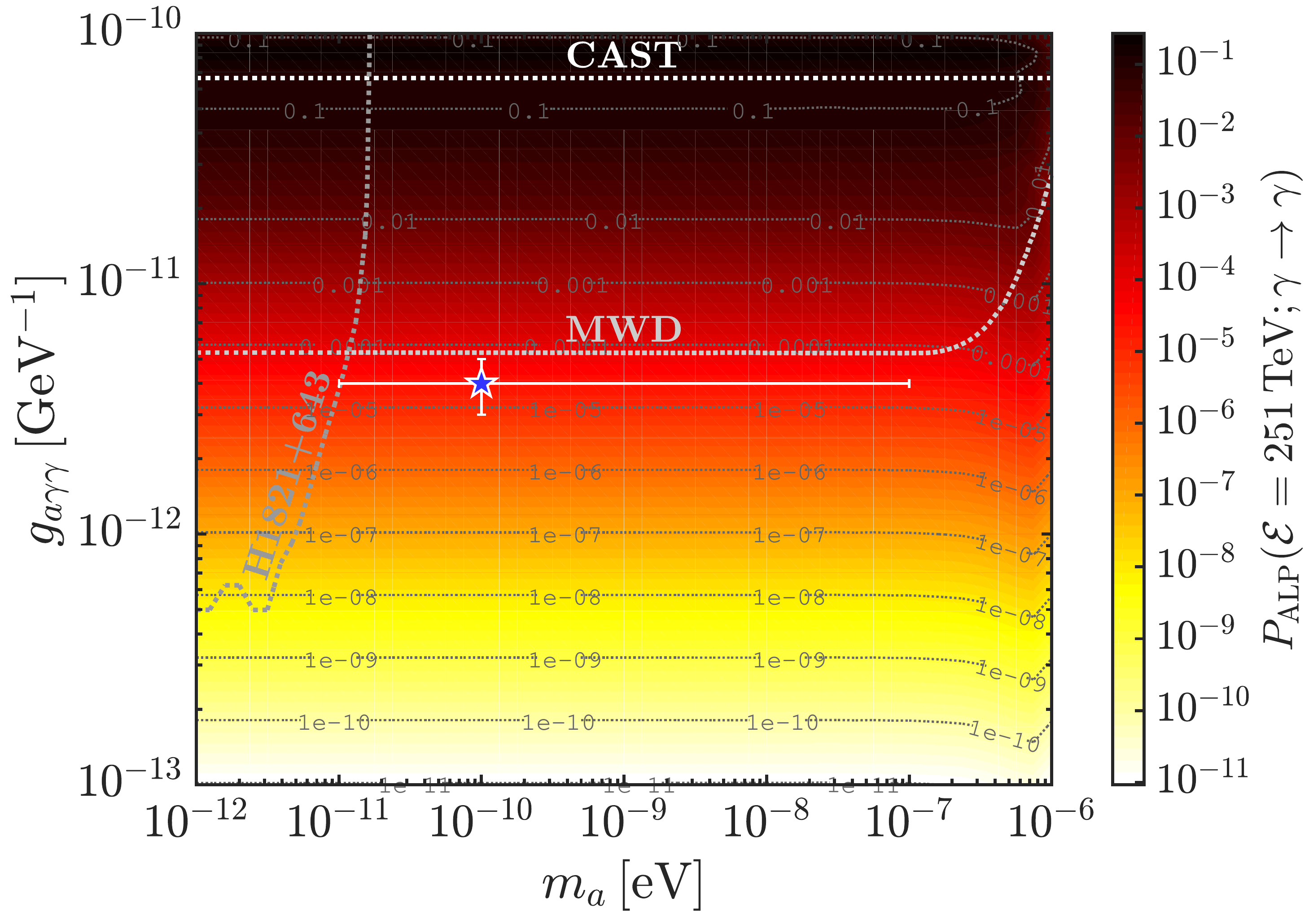}
\end{center}
\caption{\label{parSpaceStarburstBig} Photon survival probability $P_{\rm ALP}({\cal E}; \gamma \to \gamma)$ at energies ${\cal E} = 15 \, \rm TeV$ (upper-left panel), ${\cal E} = 18 \, \rm TeV$ (lower-left panel), ${\cal E} = 100 \, \rm TeV$ (upper-right panel) and ${\cal E} = 251 \, \rm TeV$ (lower-right panel), as a function of the ALP mass $m_a$ and of the photon-ALP coupling $g_{a\gamma\gamma}$ by assuming the SL EBL model, $B_{\rm ext}=1 \, \rm nG$ and a starburst hosting galaxy. The blue star with error bar represents our choice for the ALP parameter space: $m_a=10^{-10} \, \rm eV$ and $g_{a\gamma\gamma}=4 \times 10^{-12} \, \rm GeV^{-1}$. The CAST bound~\cite{cast}, that coming from magnetic white dwarf (MWD) polarization~\cite{mwd} and the one derived from H1821+643~\cite{limJulia} are also plotted.}
\end{figure*}

\begin{figure}[H]
\begin{center}
\includegraphics[width=.45\textwidth]{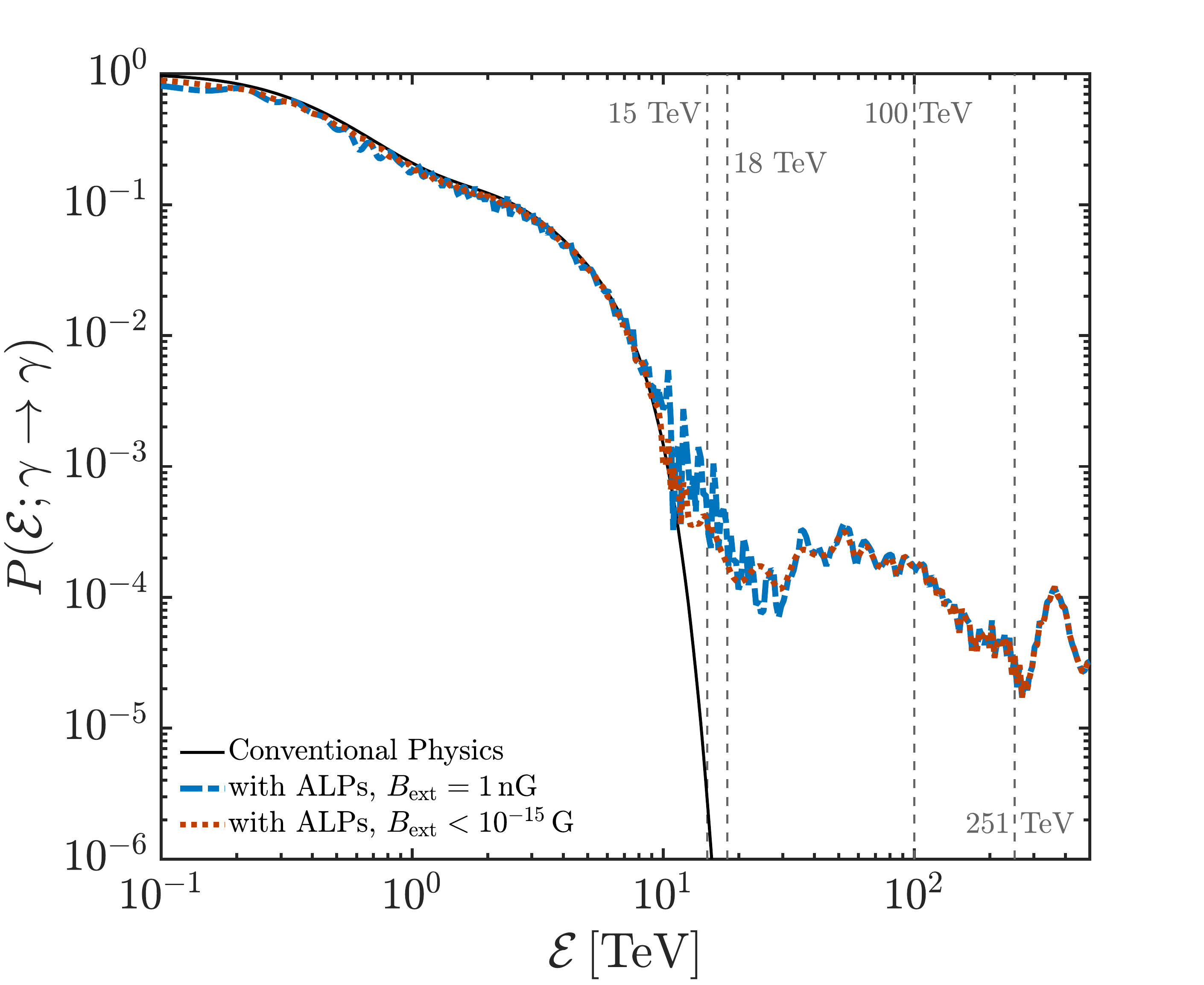}
\end{center}
\caption{\label{survProbFigStarburst} Photon survival probability $P ({\cal E}; \gamma \to \gamma)$ versus energy ${\cal E}$ in conventional physics and when ALPs (for $B_{\rm ext} = 1 \, \rm nG$ and $B_{\rm ext} < 10^{-15} \, \rm G$) are considered for the case of a starburst hosting galaxy and employing the SL EBL model. We assume $m_a=10^{-10} \, \rm eV$ and $g_{a\gamma\gamma}=4 \times 10^{-12} \, \rm GeV^{-1}$.}
\end{figure}

\begin{figure}[H]
\begin{center}
\includegraphics[width=.5\textwidth]{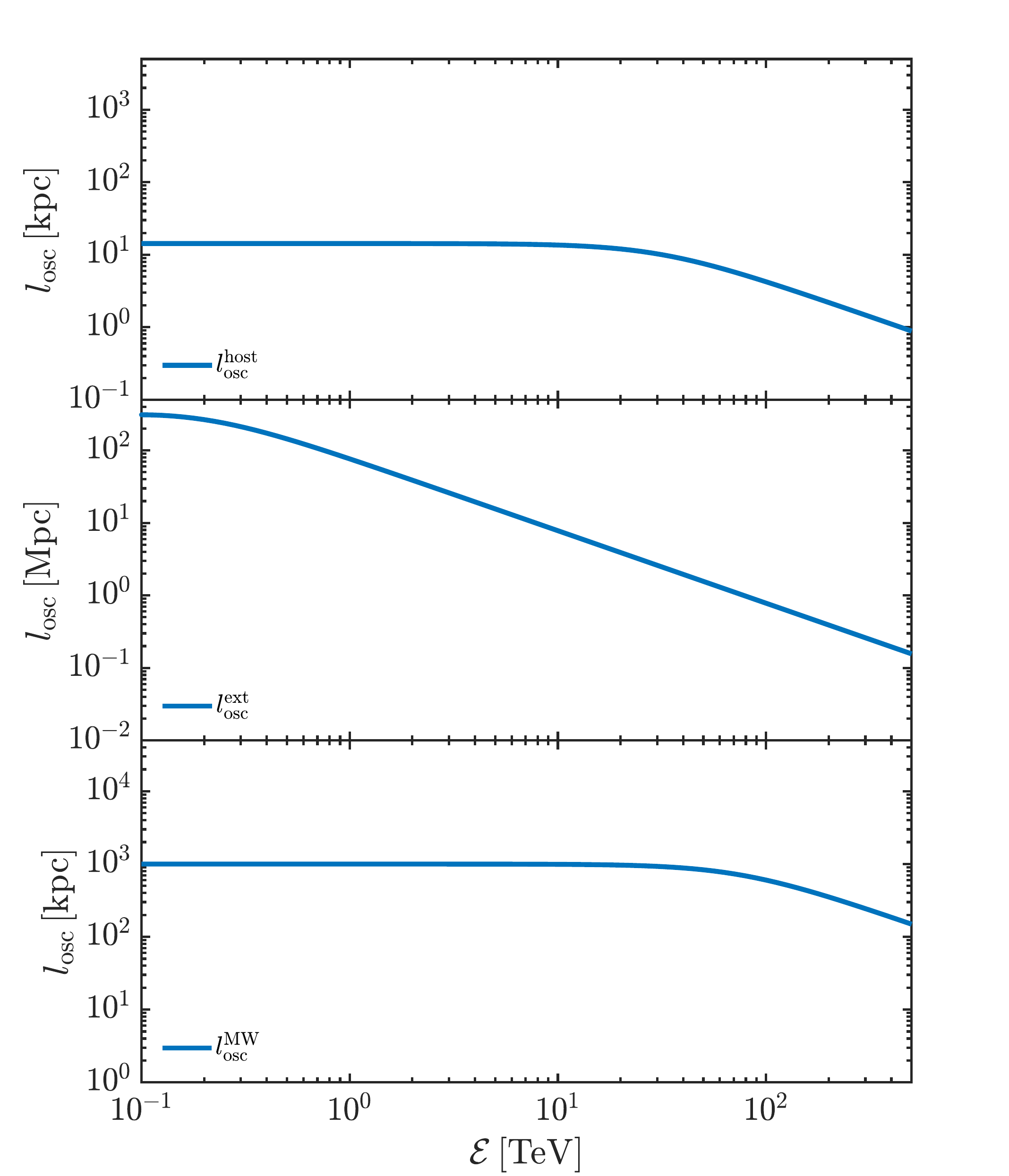}
\end{center}
\caption{\label{LoscStarburstPlot} Oscillation length $l_{\rm osc}$ versus energy ${\cal E}$ in the different crossed regions: the starburst galaxy hosting the GRB ($l_{\rm osc}^{\rm host}$, upper panel), the extragalactic space ($l_{\rm osc}^{\rm ext}$ with $B_{\rm ext} = 1 \, \rm nG$, central panel) and the Milky Way ($l_{\rm osc}^{\rm MW}$, lower panel). We assume $m_a=10^{-10} \, \rm eV$ and $g_{a\gamma\gamma}=4 \times 10^{-12} \, \rm GeV^{-1}$.}
\end{figure} 

\begin{figure}[H]
\begin{center}
\includegraphics[width=.5\textwidth]{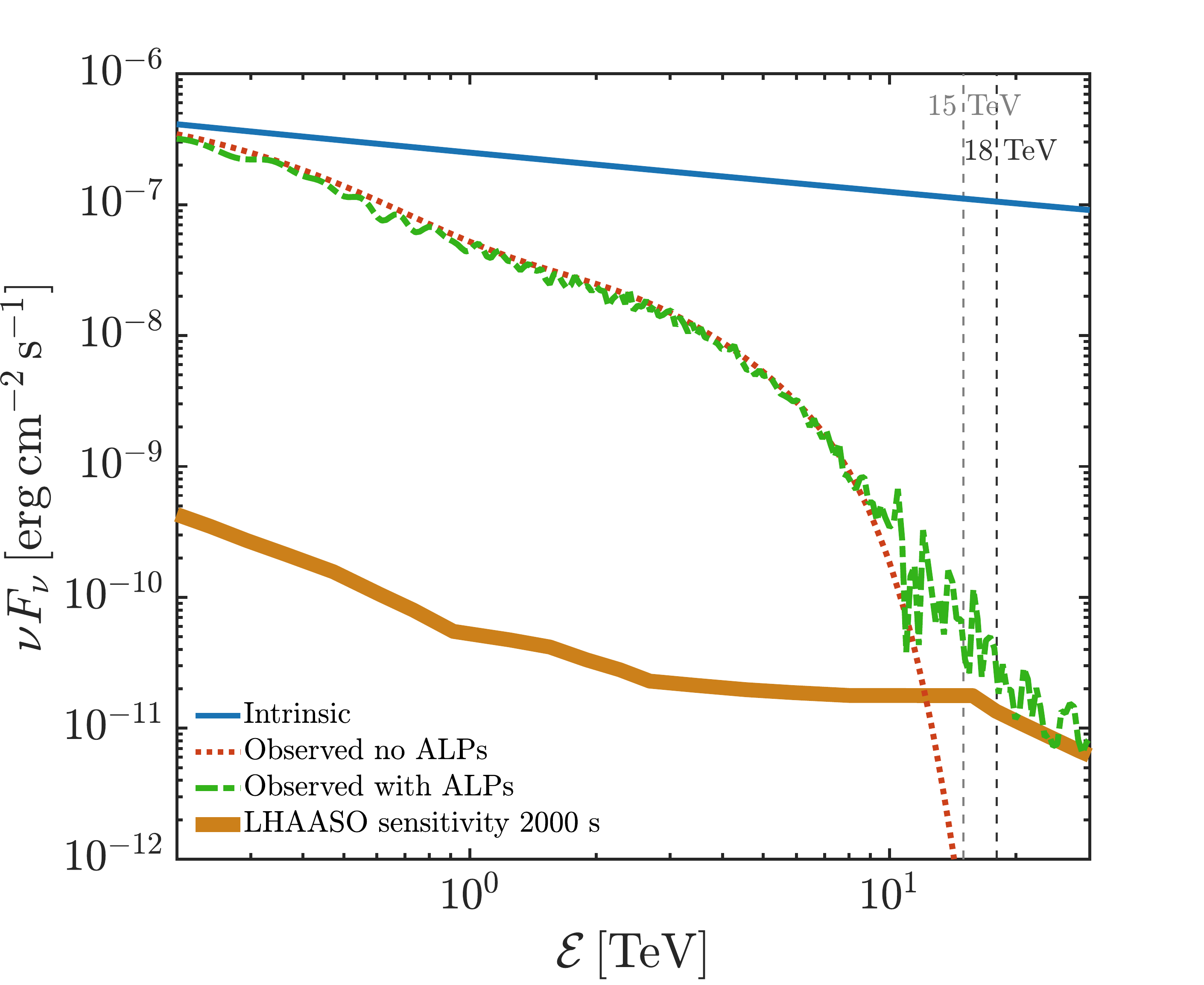}
\end{center}
\caption{\label{SpectrumStarburstBig} Intrinsic average spectrum of GRB 221009A as measured by LHAASO~\cite{grb221009aSpectrum} extended up to $\sim 20 \, \rm TeV$ and corresponding observed one versus energy ${\cal E}$ within conventional physics and when ALP effects are taken into account by assuming the SL EBL model, $B_{\rm ext}=1 \, \rm nG$, a starburst hosting galaxy, $m_a=10^{-10} \, \rm eV$ and $g_{a\gamma\gamma}=4 \times 10^{-12} \, \rm GeV^{-1}$. The LHAASO sensitivity at $2000 \, \rm s$ is also shown.}
\end{figure} 

\end{widetext}


\begin{thebibliography}{9} 

\bibitem{LHAASO} LHAASO Collaboration, GCN Circular n. 32677 (2022), \url{https://gcn.gsfc.nasa.gov/gcn3/32677.gcn3}.

\bibitem{redshift} Stargate Collaboration, GCN Circular n. 32648 (2022), \url{https://gcn.gsfc.nasa.gov/gcn3/32648.gcn3}.

\bibitem{hess} H.E.S.S. Collaboration, Science {\bf 372}, 1081 (2021).

\bibitem{carpet} Carpet-2 Collaboration, ATel \#15669 (2022), \url{https://astronomerstelegram.org/?read=15669}.

\bibitem{universe} G. Galanti and M. Roncadelli, Universe {\bf 8}, 253 (2022).

\bibitem{altri} Subsequently, a few attempts along similar lines have been put forward. See:~\cite{manuelGRB,TroitskyGRB,GonzalesGRB,CarenzaGRB,WangGRB}.

\bibitem{manuelGRB} A. Baktash, D. Horns and M. Meyer, arXiv:2210.07172 (2022).
 
\bibitem{TroitskyGRB} S. V. Troitsky, arXiv:2210.09250 (2022).
 
\bibitem{GonzalesGRB} M. M. Gonzales {\it et al.}, arXiv:2210.15857 (2022).
  
\bibitem{CarenzaGRB} P. Carenza and M. C. D. Marsh, arXiv:2211.02010 (2022).
  
\bibitem{WangGRB} L. Wang and Bo-Q. Ma, arXiv:2304.01819  (2023).

\bibitem{grb221009aSpectrum} LHAASO Collaboration, Science {\bf 380}, 1390 (2023).

\bibitem{dwek} E. Dwek and F. Krennrich, Astropart. Phys. {\bf 43}, 112 (2013).

\bibitem{SMref1} See Supplemental Material for discussion about existing EBL models, which includes Refs.~\cite{nishikov1962,gouldschreder1967,faziostecker1970,breitwheeler,heitler,hauwserdwek2001,primack2001,primack2005,gilmore2009,gilmore2012,inoue2013,frv2008,franceschinirodighiero,ciber,kneiske2002,kneiske2004,finke2010,kneiskedole2010,madaupozzetti2000,dominguez2011,schroedter2005,aharonian2006,mazinraue2007,mazingoebel2007,finkerazzaque2009,orrkrennrichdwek2011,gptr}.

\bibitem{nishikov1962} A. Nikishov, Sov. Phys. JETP {\bf 14}, 393 (1962).

\bibitem{gouldschreder1967} R. J. Gould and G. P. Schreder, Phys. Rev. {\bf 155}, 1404 (1967).

\bibitem{faziostecker1970} G. G. Fazio and F. W. Stecker, Nature (London) {\bf 226}, 135 (1970).

\bibitem{breitwheeler} G. Breit and J. A. Wheeler, Phys. Rev. {\bf 46}, 1087 (1934).

\bibitem{heitler} W. Heitler, {\it The Quantum Theory Of Radiation} (Oxford University Press, Oxford, 1960).

\bibitem{hauwserdwek2001} M. G. Hauser and E. Dwek, Annu. Rev. Astron. Astrophys. {\bf 39}, 249 (2001).

\bibitem{primack2001} J. R. Primack, R. S. Sommerville, J. S. Bullock, and J. E. G. Devriedent, AIP Conf. Proc. {\bf 558}, 463 (2001).

\bibitem{primack2005} J. R. Primack, J. S. Bullock, and R. S. Sommerville, AIP Conf. Proc. {\bf 745}, 23 (2005).

\bibitem{gilmore2009} R. C. Gilmore et al., Mon. Not. R. Astron. Soc. {\bf 399}, 1694 (2009).

\bibitem{gilmore2012} R. C. Gilmore, R. S. Sommerville, J. R. Primack, and A. Dominguez, Mon. Not. R. Astron. Soc. {\bf 422}, 3189 (2012).

\bibitem{inoue2013} Y. Inoue {\it et al.}, Astrophys. J. {\bf 768}, 197 (2013). 

\bibitem{frv2008} A. Franceschini, G. Rodighiero, and M. Vaccari, Astron. Astrophys. {\bf 487}, 837 (2008).

\bibitem{franceschinirodighiero} A. Franceschini and G. Rodighiero, Astron. Astrophys. {\bf 603}, A34 (2017). 

\bibitem{ciber} S. Matsuura {\it et al.} [CIBER Collaboration], Astrophys. J. {\bf 839}, 7 (2017). 

\bibitem{kneiske2002} T. M. Kneiske, K. Mannheim, and D. H. Hartmann, Astron. Astrophys. {\bf 386}, 1 (2002).

\bibitem{kneiske2004} T. M. Kneiske, T. Bretz, K. Mannheim, and D. H. Hartmann, Astron. Astrophys. {\bf 413}, 807 (2004).

\bibitem{finke2010} J. D. Finke, S. Razzaque, and C. D. Dermer, Astrophys. J. {\bf 712}, 238 (2010).

\bibitem{kneiskedole2010} T. M. Kneiske and H. Dole, Astron. Astrophys. {\bf 515}, A19 (2010).

\bibitem{madaupozzetti2000} P. Madau and L. Pozzetti, Mon. Not. R. Astron. Soc. {\bf 312}, L9 (2000).

\bibitem{dominguez2011} A. Dominguez {\it et al.}, Mon. Not. R. Astron. Soc. {\bf 410}, 2556  (2011).

\bibitem{schroedter2005} M. Schroedter, Astrophys. J. {\bf 628}, 617 (2005).

\bibitem{aharonian2006} F. Aharonian {\it et al.}, Astron. Astrophys. {\bf 448}, L19 (2006). 

\bibitem{mazinraue2007} D. Mazin and M. Raue, Astron. Astrophys. {\bf 471}, 439 (2007). 

\bibitem{mazingoebel2007} D. Mazin and F. Goebel, Astrophys. J. {\bf 655}, L13 (2007).

\bibitem{finkerazzaque2009} J. D. Finke and S. Razzaque, Astrophys. J. {\bf 698}, 1761 (2009).

\bibitem{orrkrennrichdwek2011} M. R. Orr, F. Krennrich, and E. Dwek, Astrophys. J. {\bf 733}, 77 (2011).

\bibitem{gptr} G. Galanti, F. Piccinini, M. Roncadelli and F. Tavecchio, Phys. Rev. D {\bf 102}, 123004 (2020).

\bibitem{saldanalopez}  A. Saldana-Lopez {\it et al.}, Mon. Not. R. Astron. Soc. {\bf 507}, 5144 (2021). 

\bibitem{astri-lhaaso} ASTRI and LHAASO Workshop, Milan, 7-8 March, 2023, \url{https://indico.ict.inaf.it/event/2288/}.

\bibitem{turok1996} N. Turok, Phys. Rev. Lett. {\bf 76}, 1015 (1996).

\bibitem{string1} E. Witten, Phys. Lett. B {\bf 149}, 351 (1984).

\bibitem{string2} J. P. Conlon, JHEP {\bf 05}, 078 (2006).

\bibitem{string3} P. Svrcek and E. Witten, JHEP {\bf 06}, 051 (2006).

\bibitem{string4} J. P. Conlon, Phys. Rev. Lett. {\bf 97}, 261802 (2006).

\bibitem{string5} K. -S. Choi, I. -W. Kim and J. E. Kim, JHEP {\bf 03}, 116 (2007).

\bibitem{axiverse} A. Arvanitaki {et al.}, Phys. Rev. D {\bf 81}, 123530 (2010).

\bibitem{abk2010} B. S. Acharya, K. Bobkov and P. Kumar, JHEP {\bf 11}, 105 (2010). 

\bibitem{cicoli2012} M. Cicoli, M. Goodsell and A. Ringwald, JHEP {\bf 10}, 146 (2012).

\bibitem{dias2014} A. G. Dias, A. C. B. Machado, C. C. Nishi, A. Ringwald and P. Vaudrevange, JHEP {\bf 06}, 037 (2014).

\bibitem{scott2017} M. J. Scott {\it et al.}, Phys. Rev. D {\bf 96}, 083510 (2017).

\bibitem{JR2010} J. Jaeckel and A. Ringwald, Ann. Rev. Nucl. Part. Sci. {\bf 60}, 405 (2010). 

\bibitem{r2012} A. Ringwald, Phys. Dark Univ. {\bf 1}, 116 (2012).

\bibitem{irastorzaredondo} I. G. Irastorza and J. Redondo, Progr. in Part. and Nucl. Phys. {\bf 102}, 89 (2018). 

\bibitem{SMref2} See Supplemental Material for details about ALPs, their astrophysical effects and existing bounds, which includes Refs.~\cite{cgn1995,straniero,fermi2016,berg,conlonLim,limFabian,limKripp,limRey2,payez2015,barblumdamico2020,meyer2020,meyerKolm,carenzaTurbB}.

\bibitem{cgn1995} S. L. Cheng, C. Q. Geng, and W. T. Ni, Phys. Rev. D {\bf 52}, 3132 (1995).

\bibitem{straniero} A. Ayala {\it et al.}, Phys. Rev. Lett. {\bf 113}, 191302 (2014).

\bibitem{fermi2016} M. Ajello {\it et al}., [Fermi-LAT collaboration], Phys. Rev. Lett. {\bf 116}, 161101 (2016).

\bibitem{berg} M. Berg {\it et al.}, Astrophys.J. {\bf 847}, 101 (2017).

\bibitem{conlonLim} J. P. Conlon {\it et al.}, JCAP {\bf 07}, 005 (2017).

\bibitem{limFabian} C. S. Reynolds {\it et al.}, Astrophys. J. {\bf 890}, 59 (2020).

\bibitem{limKripp} S. Schallmoser, S. Krippendorf, F. Chadha-Day and J. Weller, arXiv:2108.04827.

\bibitem{limRey2} J. H. Matthews {\it et al.}, arXiv:2202.08875.

\bibitem{payez2015} A. Payez {\it et al.}, JCAP {\bf 02}, 006 (2015).

\bibitem{barblumdamico2020} N. Bar, K. Blum and G. D'Amico, Phys. Rev. D {\bf 101}, 123025 (2020).

\bibitem{meyer2020} M. Meyer, T. Petrushevska and {\it Fermi} collaboration, Phys. Rev. Lett. {\bf 124}, 231101 (2020); (E) ibid. {\bf 125}, 119901 (E) (2020).

\bibitem{meyerKolm} M. Meyer, D. Montanino and J. Conrad, JCAP {\bf 9}, 003 (2014).

\bibitem{carenzaTurbB} P. Carenza {\it et al.}, Phys. Rev. D {\bf 104}, 023003 (2021).

\bibitem{rs} G. G. Raffelt and L. Stodolsky, Phys. Rev. D {\bf 37}, 1237 (1988).

\bibitem{hew1} W. Heisenberg and H. Euler, Z. Phys. {\bf 98}, 714 (1936).

\bibitem{hew2} V. S. Weisskopf, K. Dan. Vidensk. Selsk. Mat. Fys. Medd. {\bf 14}, 6 (1936).

\bibitem{hew3} J. Schwinger, Phys. Rev. {\bf 82}, 664 (1951).

\bibitem{raffelteffect} A. Dobrynina, A. Kartavtsev, and G. Raffelt, Phys. Rev. D {\bf 91}, 083003 (2015); erratum D {\bf 91}, 109902(E) (2015).

\bibitem{mpz} L. Maiani, R. Petronzio and E. Zavattini, Phys. Lett. B {\bf 175}, 359 (1986).

\bibitem{grjhea} G. Galanti and M. Roncadelli, J. High Energy Astrophys. {\bf 20}, 1 (2018).
 
\bibitem{drm2007} A. De Angelis, M. Roncadelli and O. Mansutti, Phys. Rev. D {\bf 76}, 121301 (2007).

\bibitem{simet2008} M. Simet, D. Hooper and P. D. Serpico, Phys. Rev. D {\bf 77}, 063001 (2008).

\bibitem{sanchezconde2009} M. A. S\'anchez-Conde {\it et al.}, Phys. Rev. D {\bf 79}, 123511 (2009).

\bibitem{dgr2011} A. De Angelis, G. Galanti, M. Roncadelli, Phys. Rev D {\bf 84}, 105030 (2011); (E) ibid. {\bf 87}, 109903 (E) (2013).

\bibitem{trgb2012}  F. Tavecchio, M. Roncadelli, G. Galanti and G. Bonnoli, Phys. Rev. D {\bf 86}, 085036 (2012).

\bibitem{wb2012} D. Wouters and P. Brun, Phys. Rev. D {\bf 86}, 043005 (2012).

\bibitem{trg2015} F. Tavecchio, M. Roncadelli and G. Galanti, Phys. Lett. B {\bf 744}, 375 (2015).

\bibitem{kohri2017} K. Kohri and H. Kodama, Phys. Rev. D {\bf 96}, 051701 (2017).

\bibitem{gtre2019} G. Galanti, F. Tavecchio, M. Rooncadelli and C. Evoli, Mon. Not. R. Astron. Soc. {\bf 487}, 123 (2019).

\bibitem{grdb} G. Galanti, M. Roncadelli, A. De Angelis and G. F. Bignami, Mon. Not. R. Astron. Soc. {\bf 493}, 1553 (2020).          

\bibitem{ALPpol1} P. Jain, S. Panda and S. Sarala, Phys. Rev. D {\bf 66}, 085007 (2002).

\bibitem{bassan} N. Bassan, A. Mirizzi and M. Roncadelli, JCAP {\bf 05}, 010 (2010).

\bibitem{ALPpol2} N. Agarwal, A. Kamal and P. Jain, Phys. Rev. D {\bf 83}, 065014 (2011).

\bibitem{ALPpol3} A. Payez, J. R. Cudell and D. Hutsem\'ekers, Phys. Rev. D {\bf 84}, 085029 (2011).

\bibitem{ALPpol5} R. Perna {\it et al.}, Astrophys. J. {\bf 748}, 116 (2012).

\bibitem{day} F. Day and S. Krippendorf, Galaxies {\bf 6}, 45 (2018).

\bibitem{galantiTheo} G. Galanti, Phys. Rev. D {\bf 105}, 083022 (2022).

\bibitem{galantiPol} G. Galanti, Phys. Rev. D {\bf 107}, 043006 (2023).

\bibitem{grtcClu} G. Galanti, M. Roncadelli, F. Tavecchio and E. Costa, Phys. Rev. D {\bf 107}, 103007 (2023).

\bibitem{grtBlazar} G. Galanti, M. Roncadelli and F. Tavecchio, arXiv:2301.08204.

\bibitem{magic1} V. A. Acciari {\it et al.} [MAGIC collaboration], Nature {\bf 575}, 455 (2019).

\bibitem{magic2} V. A. Acciari {\it et al.} [MAGIC collaboration], Nature {\bf 575}, 459 (2019).

\bibitem{SMref3} See Supplemental Material for details about photon-ALP propagation in GRBs, which includes Refs.~\cite{bm76,navasironi,derishevpiran,derishev}.

\bibitem{bm76} R.~D. Blandford \&  C.~F. McKee, Phys. of Fluids {\bf 19}, 1130 (1976).

\bibitem{navasironi} L. Nava, L. Sironi, G. Ghisellini, A. Celotti, G. Ghirlanda, Mon. Not. R. Astron. Soc. {\bf 433}, 2107 (2013).

\bibitem{derishevpiran} E. Derishev and T. Piran, Astrophys. J. {\bf 923}, 135 (2021).

\bibitem{derishev} E. Derishev, Mon. Not. R. Astron. Soc. {\bf 519}, 377 (2023).

\bibitem{GRB221009Ahost} A. J. Levan {\it et al}., arXiv:2302.07761.

\bibitem{SpiralBrev} R. Beck, Astron. Astrophys. Rev. {\bf 24}, 4 (2016).

\bibitem{Fletcher2010} A. Fletcher, The Dynamic Interstellar Medium: A Celebration of the Canadian Galactic Plane Survey, Astronomical Society of the Pacific Conference Series {\bf 438}, 197 (2010).

\bibitem{Thompson2006} T. A. Thompson {\it et al}., Astrophys. J. {\bf 645}, 186 (2006).

\bibitem{LopezRodriguez2021} E. Lopez-Rodriguez, J. A. Guerra, M. Asgari-Targhi and J. T. Schmelz, Astrophys. J. {\bf 914}, 24 (2021).

\bibitem{GRBposition} P. K. Blanchard, E. Berger and W. Fong, Astrophys. J. {\bf 817}, 144 (2016).

\bibitem{GRBposition2} J. D. Lyman {\it et al}., Mon. Not. R. Astron. Soc. {\bf 467}, 1795 (2017).

\bibitem{Elmegreen2004} B. G. Elmegreen and J. Scalo,  Annu. Rev. Astron. Astrophys., {\bf 42}, 211 (2004).

\bibitem{Haverkorn2008} M. Haverkorn, J. C. Brown, B. M. Gaensler and N. M. McClure-Griffiths, Astrophys. J. {\bf 680}, 362 (2008).

\bibitem{Heesen2023} V. Heesen {\it et al}., Astron. Astrophys. {\bf 669}, A8 (2023).

\bibitem{neronovvovk} A. Neronov and I. Vovk, Science {\bf 328}, 73 (2010).

\bibitem{durrerneronov} R. Durrer and A. Neronov, Astron. Astrophys. Rev. {\bf 21}, 62 (2013).

\bibitem{upbbext} M. S. Pshirkov, P. G. Tinyakov, and F. R. Urban, Phys. Rev. Lett. {\bf 116}, 191302 (2016).

\bibitem{kronberg1994} P. P. Kronberg, Rep. Prog. Phys. {\bf 57}, 325 (1994).

\bibitem{grassorubinstein2001} D. Grasso and H. R. Rubinstein, Phys. Rep. {\bf 348}, 163 (2001).

\bibitem{galantironcadelli20118prd} G. Galanti and M. Roncadelli, Phys. Rev D {\bf 98}, 043018 (2018).

\bibitem{kartavtsev} A different model has been proposed in: A. Kartavtsev,  G. Raffelt and H. Vogel, JCAP {\bf 01}, 024 (2017).

\bibitem{reessetti1968} M. J. Rees and G. Setti, Nature {\bf 219}, 127 (1968).

\bibitem{hoyle1969} F. Hoyle, Nature {\bf 223}, 936 (1969).

\bibitem{kronbergleschhopp1999} P. P. Kronberg, H. Lesch, and U. Hopp, Astrophys. J. {\bf 511}, 56 (1999).

\bibitem{furlanettoloeb2001} S. Furlanetto and A. Loeb, Astrophys. J. {\bf 556}, 619 (2001).

\bibitem{jansonfarrar1} R. Jansson and G. R. Farrar, Astrophys. J. {\bf 757}, 14 (2012). 

\bibitem{jansonfarrar2} R. Jansson and G. R. Farrar,  Astrophys. J. {\bf 761}, L11 (2012).

\bibitem{BMWturb} M. C. Beck. {\it et al.}, JCAP {\bf 05}, 056 (2016).

\bibitem{pshirkov2011} M. S. Pshirkov, P. G. Tinyakov, P. P. Kronberg and K. J. Newton-McGee, Astrophys. J. {\bf 738}, 192 (2011).                                     

\bibitem{yaomanchesterwang 2017} J. M. Yao, R. N. Manchester and N. Wang, Astrophys. J. {\bf 835}, 29 (2017). 

\bibitem{cast} V. Anastassopoulos {\it et al.} [CAST Collaboration], Nature Physics {\bf 13}, 584 (2017).

\bibitem{limJulia} J. Sisk-Reyn{\'e}s {\it et al.}, Mon. Not. R. Astron. Soc. {\bf 510}, 1264 (2022).

\bibitem{mwd} C. Dessert, D. Dunsky and B. R. Safdi, Phys. Rev. D {\bf 105}, 103034 (2022).

\bibitem{addazi} A. Addazi {\it et al.}, Progr. in Part. and Nucl. Phys. {\bf 125}, 103948 (2022).

\bibitem{gtl} G. Galanti, F. Tavecchio and M. Landoni, Mon. Not. R. Astron. Soc. {\bf 491}, 5268 (2020).

\bibitem{tavLIV} F. Tavecchio and G. Bonnoli, Astron. Astrophys. {\bf 585}, A25 (2016).

\bibitem{LIVlim}  R. G. Lang, H. Mart\'inez-Huerta, and V. de Souza, Phys. Rev. D {\bf 99}, 043015 (2019).

\bibitem{mirabal} N. Mirabal, Mon. Not. R. Astron. Soc. {\bf 519}, 85 (2023).

\bibitem{gonzalez} M. M. Gonzalez, D. Avila Rojas, A. Pratts et al., Astrophys. J. {\bf 944} 178 (2023).

\bibitem{das} S. Das and S. Razzaque, Astron. Astrophys. {\bf 670}, 12 (2023).

\bibitem{zhao} Z.-C. Zhao, Y. Zhou, and S. Wang, EPJC {\bf 83}, 92 (2023).

\bibitem{sahu} S. Sahu, B. Medina-Carrillo, G. S\'anchez-Col\'on, and S. Rajpoot, Astrophys. J., {\bf 942} 30 (2023).

\bibitem{arias2012} P. Arias {\it et al.}, JCAP {\bf 06}, 013 (2012).

\bibitem{astri} S. Vercellone {\it et al.}, J. High Energy Astrophys. {\bf 35}, 1 (2022).

\bibitem{cta} \url{https://www.cta-observatory.org/}

\bibitem{g400} A. E. Egorov {\it et al.}, JCAP {\bf 11}, 049 (2020).

\bibitem{hawc} \url{https://www.hawc-observatory.org/}

\bibitem{herd} X. Huang {\it et al.}, Astropart. Phys. {\bf 78}, 35 (2016).

\bibitem{LHAASOsens} Z. Cao {\it et al.}, Chinese Phys. C {\bf 46}, 035001 (2022).

\bibitem{desy} \url{https://taiga-experiment.info/taiga-hiscore/}

\bibitem{alps2} R. B\"ahre {\it et al.}, J. of Instrum. {\bf 8}, T09001 (2013). 

\bibitem{iaxo} I. G. Irastorza {\it et al.} [IAXO Collaboration], JCAP {\bf 06}, 013 (2011).

\bibitem{iaxo2} E. Armengaud {\it et al.}, JCAP {\bf 06}, 047 (2019).

\bibitem{stax} L. M. Capparelli {\it et al.}, Phys. Dark Univ. {\bf 12}, 37 (2016).

\bibitem{avignone1} F. T. Avignone III, Phys. Rev. D {\bf 79}, 035015 (2009).

\bibitem{avignone2} F. T. Avignone III, R. J. Crewick and S. Nussinov, Phys. Lett. B {\bf 681}, 122 (2009).

\bibitem{avignone3} F. T. Avignone III F. T., R. J. Crewick and S. Nussinov, Astropart. Phys. {\bf 34}, 640 (2011). 

\bibitem{abracadabra} Y. Kahn, B. R. Safdi and J. Thaler, Phys. Rev. Lett. {\bf 117}, 141801 (2016).

\bibitem{LHAASOspectrumHigh} Z. Cao et al., arXiv:2310.08845.

\bibitem{LHAASOarea} F. Aharonian {\it et al.}, Chinese Phys. C {\bf 45}, 025002 (2021).

\bibitem{foot1} Other processes discussed in~\cite{gptr} are totally irrelevant for the energy ranges considered in this paper.

\bibitem{remark2} It has been shown in~\cite{carenzaTurbB} that the inclusion of the turbulent component can change the photon survival probability in the Milky Way by at most a factor of 2.

\end{thebibliography}
\end{document}